\documentclass[aps,prl,reprint,floatfix]{revtex4-2}
\usepackage{graphicx,xcolor,amssymb,amsmath,hyperref,dcolumn,bm}
\usepackage[print-unity-mantissa=false]{siunitx}
\PassOptionsToPackage{normalem}{ulem}%%%%%%%
\usepackage{ulem}
\usepackage{physics}\usepackage{empheq}

\newcommand{\vect}[1]{\mathbf{#1}}

\usepackage[version=4]{mhchem}

\newcommand{\figref}[1]{Fig.~\ref{#1}}

\graphicspath{{Figures/}}

\makeatletter
\renewcommand{\frontmatter@thefootnote}{%
  \ifcase\value{mpfootnote}\or
    \@fnsymbol{2}% first footnote: dagger
  \or
    \@fnsymbol{1}% second footnote: asterisk
  \else
    \@fnsymbol{\value{mpfootnote}}%
  \fi
}
\makeatother

\begin{document}

%\title{A crawling interfacial droplet by active particles inclusion\\
\title{Active colloidal rafts perform swimming strokes that propel their host droplets}
\author{Airi N. Kato$^{1,4}$}
\thanks{These authors contributed equally to this work.}
\author{Julien Brémont$^{2,3}$}
\thanks{These authors contributed equally to this work.}
\author{Etienne Harté$^{1}$}
\author{Jean-Michel Rampnoux$^{1}$}
\author{Jean-François Joanny$^{2}$}
\author{Hamid Kellay$^{1}$}
%\email{hamid.kellay@u-bordeaux.fr}
\affiliation{
$^{1}$ Laboratoire Ondes et Mati\`ere d’Aquitaine, Universit\'e de Bordeaux, Talence 33405, France
}
\affiliation{$^{2}$Collège de France, 3 rue d'Ulm, 75005 Paris, France}
\affiliation{$^{3}$Max Planck Institute for the Physics of Complex Systems, Nöthnitzer Str. 38, 01187 Dresden, Germany}
\affiliation{$^{4}$School of Interdisciplinary Sciences, Meiji University, Tokyo 164-8525, Japan}

% \author{A. N. Kato}
% \thanks{These authors contributed equally to this work.}
% \affiliation{
% $^{1}$ Laboratoire Ondes et Mati\`ere d’Aquitaine, Universit\'e de Bordeaux, Talence 33405, France
% }
% \author{J. Brémont}
% \thanks{These authors contributed equally to this work.}
% \affiliation{Collège de France, 3 rue d'Ulm, 75005 Paris, France}
% \affiliation{Max Planck Institute for the Physics of Complex Systems, Nöthnitzer Str. 38, 01187 Dresden, Germany}
% \author{E. Harté}
% \affiliation{
% $^{1}$ Laboratoire Ondes et Mati\`ere d’Aquitaine, Universit\'e de Bordeaux, Talence 33405, France
% }
% \author{J. M. Rampnoux}
% \affiliation{
% $^{1}$ Laboratoire Ondes et Mati\`ere d’Aquitaine, Universit\'e de Bordeaux, Talence 33405, France
% }
% \author{J. F. Joanny}
% \affiliation{Collège de France, 3 rue d'Ulm, 75005 Paris, France}
% \author{H. Kellay}
% \affiliation{
% $^{1}$ Laboratoire Ondes et Mati\`ere d’Aquitaine, Universit\'e de Bordeaux, Talence 33405, France
% }%

% %\email{airi.nakamoto@u-bordeaux.fr}
% \author{Etienne Harte$^{1}$}
% \author{Kaili Xie$^{1,2}$}
% \author{Benjamin Gorin$^{1}$}
% \author{Jean-Michel Rampnoux$^{1}$}
% %\author{Alois W\"{u}rger$^{1}$}
% \author{Hamid Kellay$^{1}$}%
%  \email{hamid.kellay@u-bordeaux.fr}
% \affiliation{
% $^{1}$ Laboratoire Ondes et Mati\`ere d’Aquitaine, Universit\'e de Bordeaux, Talence 33405, France
% }%
% \affiliation{%
%  $^{2}$Van der Waals-Zeeman Institute, Institute of Physics, University of Amsterdam, 1098XH Amsterdam, The Netherlands
% }%
\begin{abstract}
   % While a single light-driven Janus particle confined in a droplet lens floating on a liquid interface displays unexpected periodic circular motions, a raft of such active particles gives rise to a nontrivial mobility of the droplet. The mechanism for the mobility of this structure (particle raft+drop) relies on the coupling between the off center raft mobility and droplet deformation. We unveil the properties of this crawling from experiments and propose a model for the mobility of the droplet which explicitly takes into account the trajectories of the rafts and the deformation of the droplet.  
    Active particles confined in droplets can set their host droplet into motion. This phenomenon is usually attributed to direct hydrodynamic forcing by the enclosed particles. Here we reveal a different mechanism: we show that a self-propelled raft of active colloids confined in a thin oil droplet at the surface of water typically follows an off-centered orbit, thus periodically deforming the droplet's contact line in an anisotropic way. This non-reciprocal deformation acts as a swimming stroke and propels the droplet according to the principles of low-Reynolds-number swimming. The effect is absent for a single colloid and arises from the intrinsically many-body dynamics of the raft, which we capture with a dynamical model coupling viscous hydrodynamics and many-body active dynamics.
    % \ank{$\leftarrow$it is also due to softer interfaces. I hesitate to say this}
    Our results show how collective internal activity can be converted into locomotion through deformation of a soft confining boundary.
\end{abstract}
\date{\today}

\maketitle

Active particles which are confined inside droplets, vesicles or membranes can deform, reshape, and possibly move these soft compartments~\cite{Sanchez2012,adkins_dynamics_2022,xie_activity_2022,kokotSpontaneousSelfpropulsionNonequilibrium2022,Sakamoto2022,Sakuta2023,le_nagard_encapsulated_2022,vutukuri_active_2020,takatori2020active,deblais_boundaries_2018,junot_active_2017,grauerActiveDroploidsa,Schonhofer2025}. Soft compartments are of interest at the fundamental level \cite{paoluzzi2016shape,reighSwimmingCageLowReynoldsnumber2017} to understand the interplay between activity and large scale deformations and possible mobility but also for practical uses such as interface cleaning,  enhanced oil recovery~\cite{Hickl2022}, biofilm formation~\cite{Hickl2022, Prasad2023}, targeted drug delivery~\cite{Luo2018} or soft robotics \cite{deblais_boundaries_2018,boudet2021collections}.

In return, confining boundaries or confining potentials actually strongly influence active particle dynamics which may show non-trivial features ~\cite{Kato2025,dauchot2019}. For example, the dynamic behavior of a single light-driven Janus particle (JP)~\cite{Kato2025} strongly confined in a very thin lens-like droplet floating at a liquid interface reveals a rich phenomenology with trajectories going from periodic regular circular orbits to irregular chaotic ones: the droplet acts as a confining potential and therefore leads to additional forces and torques acting on the particle ~\cite{Kato2025,dauchot2019}. Here, we find that when active particles assemble into mobile rafts, a new phenomenology arises whereby the confining droplet itself acquires mobility at the liquid interface. This brings forth the fundamental question of the mechanism driving droplet mobility. 
In this Letter, we show experimentally and theoretically how the active dynamics of a solid raft of JPs can make its confining droplet swim. Our central result is that the raft's dynamics typically break rotational symmetry and turn into an effective swimming stroke of the droplet, a mechanism absent in the single-JP case as studied in~\cite{Kato2025}.
%We provide an experimental and theoretical study of this mobility and of the interplay between raft trajectories and droplet velocity. 
We thereby identify a new route to active transport of a soft droplet where collective internal activity propels the confining droplet. As an application, liquid lens mobility at liquid interfaces opens new perspectives for surface cleaning, transport and delivery.
\begin{figure}[h]
\includegraphics[width=\columnwidth]{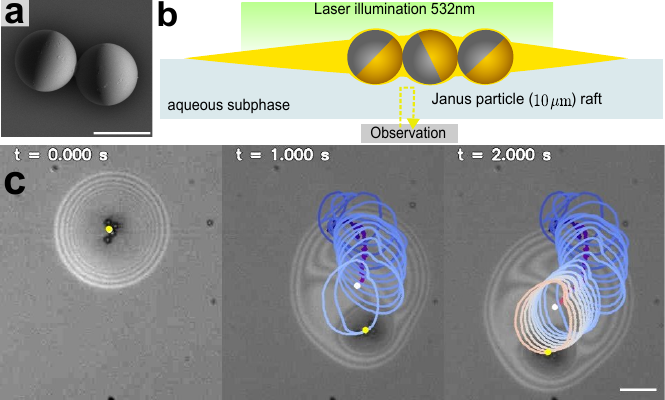}
\caption{\label{fig:1}(a) SEM micrograph of Janus particles. These polystyrene particles are half coated with a thin gold layer. Scalebar: $\SI{10}{\micro\metre}$. (b) Sketch of the experimental setup where an oil lens floats on an aqueous phase. A laser beam is used to activate the Janus particles while observations are carried out from below. (c) Montage of a four-particle raft motion in a drop at three different times. The past trajectory of the raft's center (yellow dot) orbits around the droplet's center (white dot) whose past trajectory is indicated by the thick solid line. The color change of both raft and droplet trajectories indicate elapsed time, going from darker to lighter color. Scalebar: $\SI{50}{\micro\metre}$. }
\end{figure}

The self-propelled Janus particles used in our work are polystyrene spheres, of radius $a_{\mathrm{ JP}}=\SI{5}{\micro\metre}$, half-coated with a metal layer as shown in Fig.1a~\cite{Kato2025,xie_activity_2022}. They were suspended in  dodecane oil containing 10mM of the ionic surfactant (AOT, bisethylhexyl sulfosuccinate, above the critical micellar concentration)~\cite{binks1991effects,kellay1992wetting}. A drop of this suspension was then placed onto an aqueous subphase (ultrapure water) in a Petri dish and chopped with a needle to create separated smaller lenses. Each lens encloses several JPs, which form a rigid raft residing at its center as shown in Fig.1b and c. Activity of the JP raft was induced by  a laser beam \cite{xie_activity_2022} and tuned by varying the illumination intensity $I$ of the top-hat profile laser (wavelength: \SI{532} {\nano\metre} and radius: \SI{108} {\micro\metre})~\cite{xie_activity_2022}. Initially, the center of the illumination area, which is larger than the drop size, was aligned with the center of the stationary droplet. The dynamics of the JP raft and the droplet were visualized using an inverted microscope (Axio Observer, Zeiss) equipped with a fast camera (Phantom v640 from Vision Research Inc.) working at frame rates greater than 1000 fps.
 
% \begin{figure*}[h]
% \includegraphics[width=\textwidth]{figures/RaftTrajectory.png}
% \caption{\label{fig:rafttraj} Periodic circular trajectory of a raft drawn on a few snapshots in a surfactant-laden droplet. Scalebar: $\SI{50}{\micro\metre}$. }
% \end{figure*}

The lenses were sufficiently thin to allow for the measurement of their thickness profiles using the clearly visible fringe pattern, as seen in \figref{fig:1}c. Enclosed JPs move due to thermal Marangoni flows, and a single JP was shown to exhibit two types of trajectories, either irregular or circular and periodic~\cite{Kato2025}. In neither case was droplet translation observed~\cite{Kato2025}; while the thickness profile of the confining droplet changed dynamically, the contact line remained essentially fixed in the periodic case.
Here, we show how enclosing multiple JPs, self-assembled into a rigid raft of $4$ to $6$ particles, leads to qualitatively different behavior. We additionally soften the droplet interface through the addition of surfactants. 
% \ju{Dumb question: do single JP in this softer interface case also perform centered trajectories, as in the no-surfactant case? I guess so but this is important}\ank{I have two videos of single JP. One is well-confined, the JP didn't go to close to the contact line therefore no droplet motion. Another one might be exceptional; it is in a very small (and thin) droplet. Maybe heating affect a lot, and dilation and deformation are large but no persistency for the droplet centroid. }. 

% The active raft and the droplet were well segmented from the captured image sequences using the trainable Weka segmentation~\cite{Berg2019} or Ilastik~\cite{ArgandaCarreras2017}. 
Upon laser illumination, the active raft showed circular-like periodic motion with velocities at $\mathcal{O}(\SIrange{1}{10}{\milli\metre\per\second})$ and period $T \approx \SI{0.156}{\second}$. The center of the trajectory remained distinct from the center of the drop: the trajectories were generally off-centered. Fig.~1c shows the superposition of the droplet and the raft trajectories at different instants in time.  

\begin{figure*}
\includegraphics[width=0.95\textwidth]{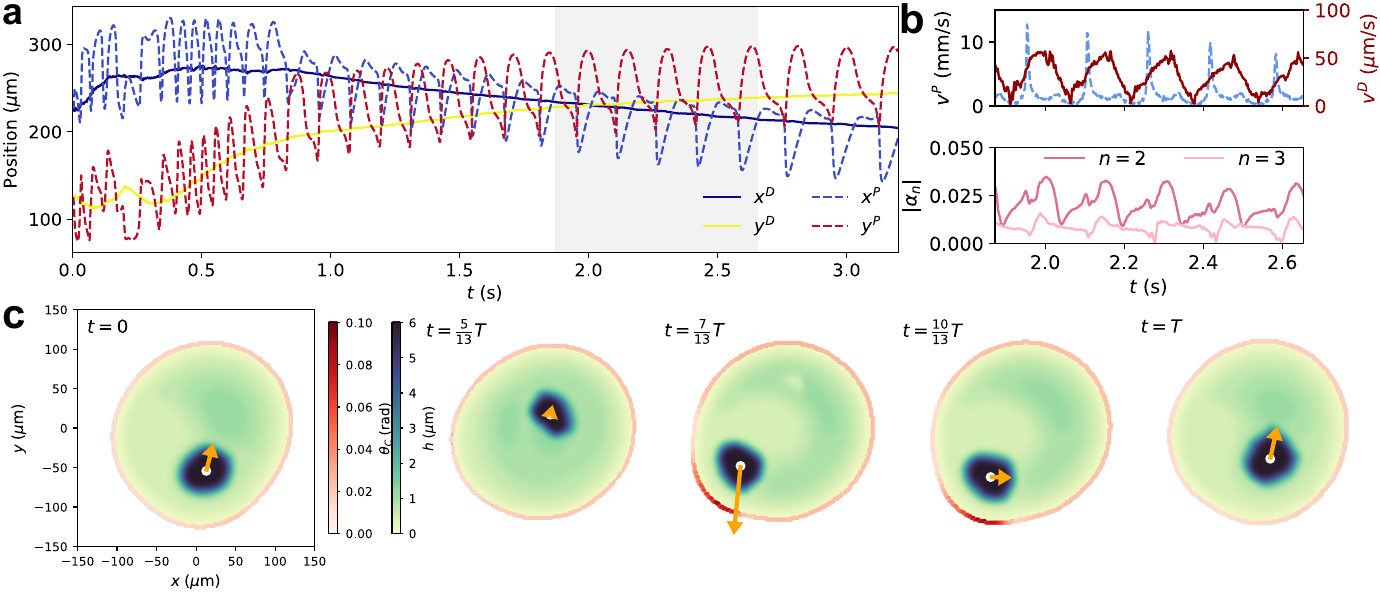}
\caption{\label{fig:2} (a) Time series of the particle raft position $(x^P,y^P)$ and the droplet position $(x^D,y^D)$. Note that the droplet moved a distance of more than $\SI{100}{\micro\metre}$ comparable to its size while the colloid raft orbits rapidly and asymmetrically around the droplet center. (b) Speeds of the particle raft $v_P$ and droplet ($v^D$, top figure) and the deformation amplitude of the droplet $|\alpha_n|$ (bottom figure) for the time window of the shaded region in (a). Note that the particle raft velocity shows sharp peaks while the droplet velocity and its deformation show broader peaks that are synchronized with the particle movement. (c) Montage of the reconstructed oil film thickness over one period starting from $t=2.185$s. The peripheral colors are coded by the local contact angle $\theta_\mathrm{C}(\varphi)$ %(a) and $\frac{\partial R}{\partial t}(\varphi)$
where $\varphi$ is the azimuthal angle. The centroid of the particle raft is marked as a white dot with the raft velocity indicated by an arrow; the length and direction of this arrow indicate the velocity magnitude and direction. Note that the contact angle increases strongly when the particle is close to the contact line as in time $7/13T$. Also note that the contact line deformation travels along the drop contour as the raft orbits. It is clear here that the raft's trajectory is biased towards the lower-left region of the droplet. }
\end{figure*}%(b) Droplet thickness reconstruction, from interference fringes, at $t=t_0:=\SI{1.873}{\second}$. The dark region is due to the presence of the particle raft. This reconstruction allows to obtain the instantaneous thickness and drop shape as well as the contact angle of the drop. 

Along with this raft motion, the confining droplet exhibited significant net motion at a velocity $\approx \SI{30}{\micro\metre\per\second}$, as shown in \figref{fig:1}c and \figref{fig:2}a (See also Supp. Movie S1). Importantly, a single JP still produces no measurable droplet translation, even for a case of a softened droplet (Supp. Movie S2). The thick line in \figref{fig:1}c shows the trajectory of the center of mass of the droplet which moves over a distance comparable to or greater than its radius; the periodic trajectory around the moving droplet center is that of the particle raft.  
The time evolution of the positions of the particle raft and the droplet are shown in \figref{fig:2}a. Note that the center of the raft trajectory is clearly distinct from the center of the drop. This observation is crucial for droplet mobility as we explain below. While the droplet moves smoothly and relatively slowly, the particle raft moves rapidly around the droplet. 

The speeds of the JP raft and the droplet change periodically in time, \figref{fig:2}b. The raft's velocity shows periodicity with spikes of several mm/s while the drop velocity changes with the same period but more smoothly with values near tens of $\SI{}{\micro\metre\per\second}$. In addition to the periodic motion of the particle and droplet center, the droplet deformation, measured by the Fourier modes of the deviation from a circle with the average droplet radius $a(t)$: $\alpha_n$ ($n \geq 2$), also varies periodically in time and seems synchronized and correlated with the motion of the droplet center of mass as shown in \figref{fig:2}b.
From images of the droplet and the observed fringes, we obtain the thickness of the drop by a reconstruction method using the fringe patterns, as shown in \figref{fig:2}c. From such images, the deformation of the oil droplet can be obtained reliably. This reconstruction actually allows us to visualize the thickness changes and also obtain the contact angle of the droplet at all instants. To better visualize the particle motions and the oil droplet three-dimensional deformation, we show the contact angle $\theta_\mathrm{C}(\varphi,t)$ along the droplet contact line using a color code in \figref{fig:2}c. It increases only when the raft comes close to the contact line.

The coupling between internal activity, droplet deformation, and
self-propulsion has recently been addressed in hydrodynamic frameworks for
active particles enclosed in droplets~\cite{huangActiveDroplet,
kawakamiMigrationDeformation,kree_controlled_2021,sprenger_towards_2020}. In
these descriptions, the confined active particle acts as a direct
hydrodynamic source, generating a flow that drives droplet migration whether
the particle remains centered or not~\cite{reighSwimmingCageLowReynoldsnumber2017}. The
mechanism at work here is qualitatively different. Droplet motion is not a
simple consequence of the activity or of the Marangoni flow driven by the
confined JPs: as said earlier, a single JP self-propels rapidly and produces a strong
Marangoni flow, yet it causes neither appreciable deformation of the contact
line nor measurable droplet translation~\cite{Kato2025}. The
new ingredient necessary for droplet motion is therefore not simply stronger
activity, but a spontaneous breaking of rotational symmetry generated by the
many-body dynamics of the rigid JP raft.
% The resulting raft trajectories (Supplemental Movie 1) are qualitatively different from the
% single-JP circular orbits (Supplemental Movie 2), forming distorted
% circles whose center does not coincide with that of the droplet, as seen in
% Fig.~1c. 
As we proceed to show, the off-centered raft trajectories constitute a swimming stroke \cite{laugaFluidDynamics,
shapereSelfPropulsionLow,stonePropulsionMicroorganisms} via non-reciprocal deformation of the contact line by the raft.

\begin{figure}[h]
\includegraphics[width=\columnwidth]{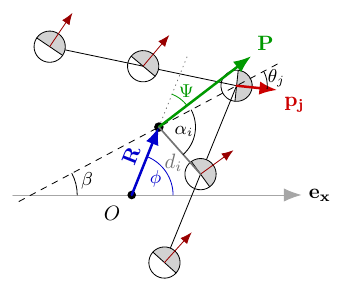}
\includegraphics[width=\columnwidth]{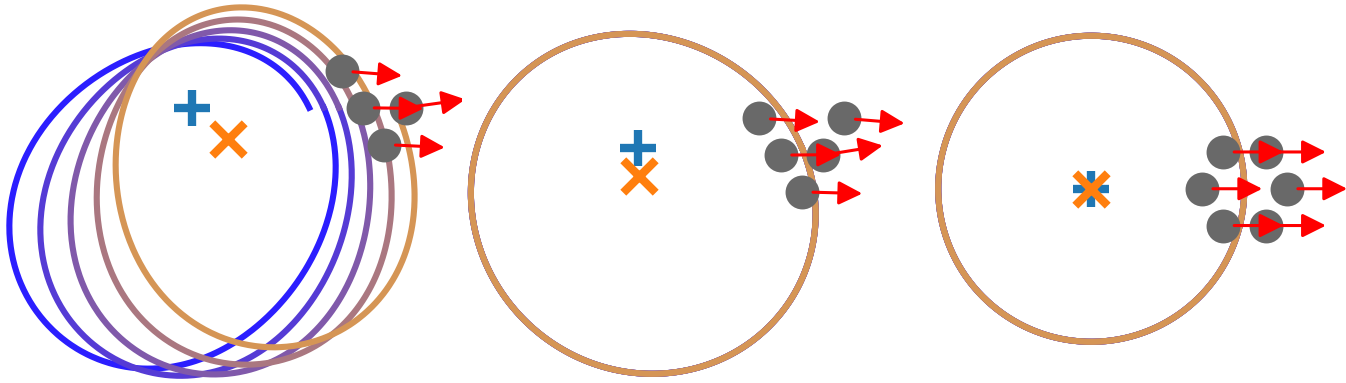}
\caption{\label{fig:raft-geometry} Top: Visual representation of the raft's geometry and JP polarities (here, for a V-shaped raft of $N=5$ JPs) and of our geometric notations. $O$ is taken to be the droplet's center. Bottom: Trajectories simulated using our minimal model Eq.~\eqref{eq-dynamic} for various raft geometries (grey circles) and polarities (red arrows). Blue orbits occur earlier than yellow ones. The blue cross represents the droplet's center while the orange one is the last orbit's center. Observe that the V-shaped raft with $N=4$ JPs and slight polarity disorder exhibits trajectories qualitatively similar to that of Fig.~\ref{fig:1} in the same setting, while a $N=6$ hexagon with no polarity disorder $\delta=0$ moves in a perfect circle, exactly like the $N=1$ single JP case \cite{Kato2025}. Finally, a $N=5$ V-shaped raft acquires a stable off-centered orbit that does not precess with time. See the SM for more details.}
\end{figure}

We first explain the observed raft trajectories by treating the raft as a rigid cluster of $N$ JPs as shown in Fig.~\ref{fig:raft-geometry}: experimentally, we observe that pairwise distances between the JPs stay constant. We denote by $\mathbf{R}=R\mathbf{e}(\phi)$ the position of the raft's center of mass relative to the droplet center $O$, where $\mathbf{e}(\phi)$ is a unit vector in the direction $\phi$, and by $\beta$ the orientation of the solid raft in the laboratory frame. The position of particle $i$ is then $\mathbf{r}_i=\mathbf{R}+d_i \mathbf{e}(\beta+\alpha_i)$,
where the lengths $d_i$ and angles $\alpha_i$ encode the static raft geometry. Each JP has a polarity $\mathbf{p}_i=\mathbf{e}(\beta+\theta_i)$, pointing toward its metallic cap. These define the mean polarity $\mathbf{P}\equiv \frac{1}{N}\sum_{i=1}^N \mathbf{p}_i
=P_0 \mathbf{e}(\phi+\Psi),$
where $\Psi$ is the angle between $\mathbf{P}$ and $\mathbf{R}$ and $0\leq P_0\leq 1$.
% \ank{Do we write it explicitly $P_0=\vert \frac{1}{N}\Sigma_i^N e^{i\theta_i}\vert =\vert \frac{1}{N}\Sigma_i^N e^{i\delta_i}\vert $ by choosing $\theta$ to the direction of mean polarity. Then $\delta=\arg (\frac{1}{N}\Sigma_i^N e^{i\delta_i})$, same as the latter eq. with $P_0$.} \ju{after some thought I prefer it this way, because without mentioning friction, $\delta$ depends on time... so it becomes less clear why we do this.}

The droplet acts on the raft in two ways. First, capillary confinement by the curved interface is modelled by an effective restoring force $-k\mathbf{r}_i$ on each particle, with a stiffness $k$ set by the interfacial tension and droplet geometry \cite{Kato2025} (our analysis applies to generic confining potentials, see Supplementary Material (SM) \cite{SM}). Second, laser heating of the metallic caps creates a local temperature gradient along the droplet interface. This gradient drives Marangoni flows and propels each particle with force $\xi v_p\mathbf{p}_i$, where $v_p$ is the self-propulsion speed and $\xi$ an effective translational drag. Since the particles form a compact raft, temperature gradients are controlled by the collective polarity $\mathbf{P}$. 
% For a single JP, a torque acting on the particle was found to be necessary to explain the trajectories observed under laser heating \cite{Kato2025}. 
The simplest expression allowed by symmetry for the capillary torque on JP $i$ is
\[
\tau_i=\Gamma(\mathbf{r}_i\times\mathbf{P})\cdot\mathbf{e}_z,
\]
where $\Gamma$ is a phenomenological coefficient.
This torque is odd under polarity reversal, increases with the distance from the droplet center, where confinement by the lens is stronger, and vanishes when the collective polarity $\mathbf{P}$, the main cause of surface tension gradients close to the raft, is radial. For a single JP, it reduces to the form found in \cite{Kato2025}, thus requiring $\Gamma>0$.

We now assume that interparticle friction rapidly locks the relative angular velocities of the JPs.  After a short transient, the polarity angles relative to the solid raft can be written as $\theta_i(t)=\theta(t)+\delta_i$, where the constants $\delta_i$ account for initial polarity disorder.
% (\textcolor{red}{have we defined $\theta?$ and P appears twice with different angles....is this necessary in the main text????}).
% \ank{maybe $\theta$ was not defined yet? } \ju{better to define theta now IMO}
Hence, the angle $\Psi=\beta+\delta+\theta-\phi$ where $\delta$ is defined by $P_0 \mathbf{e}(\delta) = \frac{1}{N} \sum_{i=1}^N \mathbf{e}(\delta_i)$.
% \[
% \mathbf{P}
% =
% \frac{1}{N}\sum_{i=1}^N\mathbf{e}(\beta+\delta_i+\theta)
% \equiv
% P_0 \mathbf{e}(\beta+\delta+\theta).
% \]
This locking assumption is supported experimentally: in the absence of synchronization of the angular velocities, the global polarity $\mathbf{P}$ would rapidly decay as a sum of $N$ oscillations of asynchronous frequencies $\dot{\theta}_i$, whereas it is observed that the raft maintains a persistent propulsion. 

We show in SM that raft geometry and polarity disorder enter only through
\[
\sum_{i=1}^N d_i \mathbf{e}(\delta_i-\alpha_i)
\equiv
\overline d \mathbf{e}(-\overline\alpha),
\qquad
I\equiv \sum_{i=1}^N d_i^2.
\] We non-dimensionalize time and distance as $\tau=\frac{k}{\xi}t$ and $\rho=\frac{kR}{\xi v_p P_0}$
and introduce the two dimensionless control parameters
$\eta=\frac{v_p\overline d \xi}{Ik(1-c)}$ and $\mu=\frac{\Gamma v_p P_0^2\xi^2}{k^2\xi_r}$,
where $\xi_r$ is an effective rotational drag \cite{Kato2025} and $c<1$ characterizes confinement and vanishes in the infinite droplet limit (see SM). The parameter $\eta$ quantifies the coupling between raft geometry and polarity disorder; in particular, $\eta=0$ for a single JP or for a raft whose individual polarities are exactly aligned, as $\sum_{i=1}^N d_i \mathbf{e}(\alpha_i) = \mathbf{0}$ by definition of the center of mass. In turn, the parameter $\mu$ measures the strength of the active torque responsible for the raft's rotation relative to capillary relaxation, which pulls the raft towards the center of the droplet. 

The dynamics of the raft are obtained by enforcing that the total torque and force on each JP vanish in the low-Reynolds number limit. It reduces to:
\begin{align}
\dot{\rho}
&=
\cos\Psi-\rho, \quad \dot{\phi}=\frac{\sin\Psi}{\rho}, \quad \dot \beta = \eta \sin(\theta-\overline{\alpha}), \nonumber\\
\
\dot{\Psi}
&=
c\eta\sin(\theta-\overline{\alpha})
+
\left(
\mu\rho-\frac{1}{\rho}
\right)
\sin\Psi,\nonumber\\
\
\dot{\theta}
&=
\mu\rho\sin\Psi-(1-c)\eta \sin(\theta-\overline{\alpha}),
\label{eq-dynamic}
\end{align}
where the dots indicate derivative with respect to $\tau$. 
% Note that it suffices to solve Eqs.~\eqref{eq-dynamic} for the variables $(\rho, \Psi, \theta)$, as one can deduce $\beta, \phi$ from them. 
In the SM, we show that for $\mu > 1$ and to first order in $\eta/\mu$, the long-time solutions of \eqref{eq-dynamic} indeed correspond to off-centered circles :
\[
\rho(\phi)
=
\frac{1}{\sqrt{\mu}}
+
\frac{c\eta}{\mu}\cos(\phi-\phi_1)
+
O\left((\eta/\mu)^2\right).
\]
% The displacement $\ell = \tfrac{c \eta}{\mu}$ above can be understood physically. For a circular trajectory, $\dot \Psi =0$. For a small displacement $\ell$, a term $\delta \dot \Psi \propto c \eta$ appears, which balances the additional capillary torque $\Gamma \mathbf{R}\times \mathbf{P}\propto \mu \ell$.
In dimensional units, the raft trajectory at this order thus follows a circle of radius $R_0=\sqrt{\frac{v_p \xi_r}{\Gamma}}$, whose center is displaced a distance % yes dimensionally
\begin{equation}
    D_{\text{off}} \equiv \frac{c \eta}{\mu} \frac{P_0 \xi v_p}{k}= \frac{c \overline{d} v_p \xi_r}{(1-c) I P_0 \Gamma}
    \label{offcenter-formula}
\end{equation}
from the origin. This is one of our main results: the breaking of rotational symmetry required for droplet movement arises naturally from both disordered individual JP polarities and finite droplet confinement. Accordingly, $D_{\rm off}$ vanishes either when the raft is sufficiently symmetric $\overline d=0$ (e.g. when polarities are identical), or in the infinite-droplet limit, $c\to0$.  In the SM, we show numerically that this conclusion still holds at finite values of the model's parameters and for more general confining potentials. 
% As a sidenote, Eq.~\eqref{offcenter-formula} is independent of both the confinement coefficient $k$ and the translational drag $\xi$.
Finally, precession of the raft's orbit can only occur very slowly (see Fig.~\ref{fig:raft-geometry}). Indeed, the orbital angle changes over one period by $\Delta\phi=2\pi+O\left((\eta/\mu)^2\right)$, so there is no precession of the trajectory at first order and rotational symmetry cannot be restored at small times: this is consistent with Fig.~\ref{fig:1}, see also Supp. Movie S3. 
Our minimal model thus makes explicit how the circular, single-JP motion, implying $\eta=0$ \cite{Kato2025}, becomes distorted and off-centered for a raft of interacting JPs, and how this feature is inherent to both raft geometry and weak polarity disorder. 
% Finally, it also addresses the phase difference between droplet and raft velocities observed in Fig.~\ref{fig:2}, where we see that droplet velocity is highest when its deformation (measured via $\alpha_2$) is largest (i.e. the raft is closest to the contact line) and the raft's speed is the lowest. This is consistent with our minimal model \jf{for raft dynamics} which predicts a timelag of roughly one fourth of a period, see SM. 

%We are now in a position to explain droplet migration. 

Because the raft orbit is off-centered, the raft approaches the contact line preferentially along one region of the droplet. Since the local contact angle increases strongly when the raft comes close to the boundary, this produces a localized deformation that travels along the droplet perimeter as the raft moves. In the event shown in Fig.~\ref{fig:2}, for instance, the raft trajectory is biased toward the lower-left part of the droplet; from the droplet frame, the boundary is therefore driven by a traveling bump localized in that region. We now show that such a traveling boundary deformation is a swimming stroke in the sense of low-Reynolds-number swimming \cite{laugaFluidDynamics,shapereSelfPropulsionLow,stonePropulsionMicroorganisms,ishimoto_squirmer_2013,mathijssenHydrodynamicsMicroswimmersFilms2016,fortune_biophysical_2024}.

We describe the droplet's oil-water-air contact line in polar coordinates relative to the droplet's center $O$ as
\begin{equation*}
    r(\theta,t)
    =
    a\left[1+\varepsilon \xi(\theta,t)\right],
    \qquad
    \xi(\theta,t)
    =
    \sum_{|n|\neq  1}\alpha_n(t)e^{in\theta}.
\end{equation*}
Here, $a$ is the average droplet radius and $\varepsilon$ is the deformation amplitude, assumed here to be small. The coefficients $\alpha_n(t)$ are the shape modes of the contact line. Because the deformation is real, one has $\overline{\alpha_n(t)} = \alpha_{-n}(t)$, while the absence of $|n|=1$ modes fixes the droplet's center at $O$.

\begin{figure}[h]
\includegraphics[width=\columnwidth]{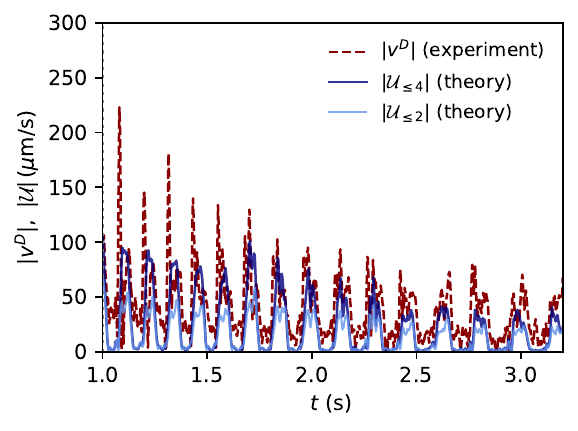}
\caption{\label{fig:comparison} Comparison between the magnitude of the droplet velocity from centroid tracking and the deformation-based prediction of the velocity, using Eq.~\eqref{velocity-final} for theory. $\mathcal{U}_{\leq n}$ means that we consider only up to terms of index $n$ in the sum Eq.~\eqref{velocity-final}. See SM for a component-wise comparison.} 
% (b) Correlation between the time-averaged speeds of the experiment and theory from eq. \eqref{velocity-final}. \ju{A reecrire, plus de components. Also, I now wonder if panel b is convincing: is it really meaningful to make a linear regression, when the magnitude $|U| = |U_{geom} + U_{translation}| \neq |U_{geom}| + |U_{translation}| $ ?... This may be the good place to put the components-wise regression $U_{geom, x/y}$ vs $U_{measured, x/y}$ since this would be linear, and to show explicitly the difference $U_{geom, x/y} - U_{measured, x/y}$ which, in the best case scenario, would be slowly varying drift. Also try plotting just the $\alpha_2 \dot \alpha_3^*$ contribution.}
\end{figure}

Given the depth of water in the Petri dish ($\approx \SI{10}{\milli\metre}$) compared to the droplet radius ($\lesssim \SI{100}{\micro\meter}$), we assume water fills the half-space below the droplet. We then isolate the contribution of the contact line's normal velocity, which is accessible experimentally unlike its tangential velocity. 
% \textcolor{red}{I don't understand what follows!!!}
% Indeed, its tangential velocity cannot be estimated reliably, as it depends on surfactants on the interface that are difficult to estimate. 
Owing to the linearity of the Stokes equations, the total droplet velocity can nevertheless be written as
\begin{equation*}
    \mathbf{U} = \mathbf{U}_{\rm geom} + \mathbf{U}_{\rm rem},
\end{equation*}
where $\mathbf{U}_{\rm rem}$ includes contributions from e.g. tangential velocities, or immeasurable deformations of the droplet surface immersed in water, which we model as a half-sphere. 
Define $\mathbf{U}=U_x+iU_y$. 
We show in the SM that
\begin{equation}
    \mathbf{U}_{\rm geom} 
    \approx
    -a\varepsilon^2
    \sum_{n\geq2}
    \left[
    A_n\alpha_n\overline{\dot\alpha_{n+1}}
    +
    B_n\overline{\alpha_{n+1}}\dot\alpha_n
    \right].
    \label{velocity-final}
\end{equation}
The speed scales quadratically with deformation amplitude $\varepsilon$ and linearly with raft angular velocity $\omega$ (via $\dot \alpha_n$), and results from the coupling of neighboring modes, in accordance with known results \cite{laugaFluidDynamics,shapereSelfPropulsionLow,stonePropulsionMicroorganisms,mathijssenHydrodynamicsMicroswimmersFilms2016,ishimoto_squirmer_2013,fortune_biophysical_2024}. 
The experimental comparison with \eqref{velocity-final} is shown in \figref{fig:comparison}. Using the experimentally acquired contact line deformation, we compute the droplet velocity Eq.~\eqref{velocity-final} without fitting parameters, then compare it directly with the measured motion of the droplet center Fig.~\ref{fig:comparison}. The calculation captures the main features of the motion, including its periodicity and the phase of the velocity variations and its order of magnitude, given by the nontrivial scale separation \[\frac{|\mathbf{U}_{\rm droplet}|}{|\mathbf{U}_{\rm raft}|} \approx \frac{a \varepsilon^2 \omega}{R_{\rm orb} \omega} \approx \varepsilon^2 \]
between droplet and raft velocities \footnote{The values of $A_n, B_n$ under a spherical approximation of the droplet given in SM show that the sum in Eq.~\eqref{velocity-final} has complex modulus of order $1$. Importantly, this sum does not introduce another order of magnitude.}.
% \ju{needs a quantitative measurement of $\varepsilon$ and the raft orbit's radius $R_{\rm orb}$}.
% In turn, $\mathbf{U}_{\rm rem}$ is, too, of the same order of magnitude as $\mathbf{U}_{\rm geom}$: this prevents us from quantitative agreement with e.g. velocity direction.  
% Quantitative differences remain, which suggest the presence of an additional residual drift contributing to droplet velocity, which is not captured by our theory based on purely normal flow velocity. A possible origin for this drift \jf{could be} an asymmetric surfactant distribution on the droplet, caused by the raft repeatedly approaching the contact line in approximately the same region of the droplet. 
% If the relaxation of this distribution is slow compared with the raft's period, a resulting surface tension gradient might persist and drive a tangential interfacial flow even when droplet deformation is weak. Such a contribution is difficult to evaluate quantitatively, since it depends on surfactant and thermal gradients at the interface and is therefore not included in our purely geometric calculation. 
%Nevertheless, its presence is consistent with our observations : this residual drift has an approximately fixed direction, decreases as the droplet deformation weakens, because surfactant distribution relaxes to become roughly uniform. Finally, it was shown that such tangential velocity was absent in the single-JP case \cite{Kato2025}, where no persistent asymmetric raft trajectory exists.

To conclude, we showed experimentally and theoretically that the coupling between the internal dynamics of active Janus particles, self-assembled into a raft, and the deformation modes of a lens droplet, can give rise to a novel swimming mode on a liquid surface. In particular, our framework explains how Marangoni driven self-propulsion, together with the raft's solid geometry, lead to asymmetric time-irreversible droplet deformations resulting in swimming of the confining droplet.

We acknowledge the support of the French Agence
Nationale de la Recherche (ANR), under Grant No. ANR-22-CE06-0007-02. We thank J. C. Baret, A. Colin, X. Wang, K. Xie, and B. Gorin for helpful and interesting discussions.

\bibliographystyle{apsrev4-2}
\bibliography{main.bib}
%\input{main.bbl}
% \appendix
% \section*{End matter}
% \smallskip

\clearpage 
\onecolumngrid
\section{Supplementary Material}
\section{Supplementary Movies}
\begin{itemize}
    \item {\bf Supplemental Movie 1} \url{S1.mp4} Periodic, off-centered circular motion of raft with the host droplet motion. x1/40 speed play. (experiment).%$I=?\SI{}{\watt\per{\centi\metre}^2}$
    \item {\bf Supplemental Movie 2} \url{S2.mp4} Periodic circular raft motion with no translation in the host droplet. x1/40 speed play. (experiment). Note that this system was with surfactants, unlike \cite{Kato2025}.
    \item {\bf Supplemental Movie 3} \url{S3.mp4} Periodic, off-centered circular motion of a V-raft of $N=5$ colloids (simulation).
\end{itemize}

\section{Computation of the droplet's swimming speed and angular rotation}
\subsection{Image construction and swimming velocity of an equatorially
deforming sphere}
\label{sec:reflected-sphere}

We estimate the velocity generated by the measured normal motion of the
contact line through an analytically tractable three-dimensional model.
The aqueous phase is treated as occupying the half-space $z<0$, while the
oil--water interface is approximated by the lower hemisphere of a sphere of
radius $a$ whose equator lies in the plane $z=0$. The upper fluid is assumed
to have negligible viscosity. Away from the droplet, the flat water--air
interface therefore obeys the impermeability and shear-free conditions
\begin{equation}
    u_z=0,
    \qquad
    \sigma_{xz}=\sigma_{yz}=0,
    \qquad z=0.
    \label{eq:free-surface-bc}
\end{equation}

\paragraph*{Reflection of the half-space problem.}
The half-space problem can be transformed into an unbounded-fluid problem
by reflecting the velocity and pressure fields across $z=0$. Denoting by
$(\mathbf u^{-},p^{-})$ the solution in $z<0$, we define in $z>0$
\begin{equation}
\begin{split}
    u_x^{+}(x,y,z)&=u_x^{-}(x,y,-z),\\
    u_y^{+}(x,y,z)&=u_y^{-}(x,y,-z),\\
    u_z^{+}(x,y,z)&=-u_z^{-}(x,y,-z),\\
    p^{+}(x,y,z)&=p^{-}(x,y,-z).
\end{split}
\label{eq:reflection}
\end{equation}
The tangential velocity components and pressure are thus even under
reflection, whereas the normal velocity is odd. Equation
\eqref{eq:reflection} preserves the Stokes equations and automatically
enforces \eqref{eq:free-surface-bc}. The mirror image of the submerged
hemisphere is an upper hemisphere, so that the reflected problem is that of
a complete deforming sphere in an unbounded fluid. Note that this same mirror construction works for a non-spherical droplet, but we use the spherical approximation in the following.

For a translation parallel to the plane, the hydrodynamic force on the
reflected sphere is twice that exerted on the submerged hemisphere,
\begin{equation}
    \mathbf F_{\parallel}^{\rm full}
    =
    2\mathbf F_{\parallel}^{\rm half}.
\end{equation}
Consequently, the force-free condition is identical in the half-space and
reflected full-space problems, and the in-plane swimming velocities are the
same.

We introduce spherical coordinates $(r,\vartheta,\varphi)$, where
$\vartheta$ is the polar angle measured from the positive $z$ axis. The
contact line corresponds to the equator $\vartheta=\pi/2$. We choose to write its
deformation as
\begin{equation}
    \xi(\varphi,t)
    =
    \sum_{n\geq2}
    \left[
        \alpha_n(t)e^{in\varphi}
        +
        \overline{\alpha_n(t)}e^{-in\varphi}
    \right].
    \label{eq:contact-line-modes-sm}
\end{equation}
We extend each azimuthal mode over the reflected sphere according to
\begin{equation}
    f(\vartheta,\varphi,t)
    =
    \sum_{n\geq2}
    \left[
        \alpha_n(t)F_n(\vartheta,\varphi)
        +
        \overline{\alpha_n(t)}
        \,\overline{F_n(\vartheta,\varphi)}
    \right],
    \qquad
    F_n(\vartheta,\varphi)
    =
    \sin^n\vartheta\,e^{in\varphi}.
    \label{eq:sectoral-extension}
\end{equation}
Since $F_n(\pi/2,\varphi)=e^{in\varphi}$, this (arbitrary, but illustrative !) choice of deformation for the entire droplet agrees
with the measured contact-line deformation \emph{only}. Note that we deliberately only take care of the contact line, and not of the entire surface in contact with water upon which the droplet sits. We chose the functions $F_n$ for simplicity as they are essentially $\ell = m = n$ spherical harmonics,
$F_n\propto \mathcal{Y}_n^n$, and vanish faster and faster as $n$ increases outside of the equator, which models the contact line (see Fig.~\ref{fig:deform-example} for an illustration).

The instantaneous surface is parametrized as
\begin{equation}
    r_s(\vartheta,\varphi,t)
    =
    a\left[1+\varepsilon f(\vartheta,\varphi,t)\right],
    \qquad \varepsilon\ll1.
    \label{eq:deformed-sphere}
\end{equation}
An isotropic correction of order $\varepsilon^2$ can be added, but it does not contribute
to translation at the order considered below due to the absence of $|n|=1$ modes.

\paragraph*{Pullback onto the undeformed sphere.}
We prescribe purely normal motion on the instantaneous deformed surface.
Introducing the angular gradient
\begin{equation}
    \nabla_\Omega
    =
    \mathbf e_\vartheta\partial_\vartheta
    +
    \frac{\mathbf e_\varphi}{\sin\vartheta}
    \partial_\varphi,
\end{equation}
the unit normal to \eqref{eq:deformed-sphere} is
\begin{equation}
    \mathbf n
    =
    \mathbf e_r-\varepsilon\nabla_\Omega f
    +O(\varepsilon^2).
\end{equation}
The deformation velocity on the deformed sphere is therefore
\begin{equation}
    \mathbf v(r_s)
    =
    a\varepsilon\dot f\,\mathbf e_r
    -
    a\varepsilon^2\dot f\,\nabla_\Omega f
    +O(\varepsilon^3).
    \label{eq:normal-velocity-deformed}
\end{equation}

We now need to pull back this expression to the undeformed sphere. We thus expand the fluid velocity and swimming velocity in powers of
$\varepsilon$ as 
\begin{equation}
    \mathbf{u} = \varepsilon\mathbf{u}_1 + \varepsilon^2\mathbf{u}_2 + \dots
\end{equation}

 Since \eqref{eq:sectoral-extension} contains no $\ell=1$
mode, the first-order translation vanishes. At first order, thanks to the no-slip condition on the deformed sphere, the boundary
condition on the undeformed sphere is purely radial
\begin{equation}
    \mathbf u^{(1)}(a)
    =
    a\dot f\,\mathbf e_r.
    \label{eq:first-order-bc}
\end{equation}
At second order, pulling back \eqref{eq:normal-velocity-deformed} to the undeformed sphere requires pulling back $a \dot f \mathbf{e_r}$: this gives
\begin{equation}
    \mathbf v_{\rm eff}^{(2)}(r=a)
    =
    -af\left.\partial_r\mathbf u^{(1)}\right|_{r=a}
    -
    a\dot f\,\nabla_\Omega f.
    \label{eq:effective-second-order-gait}
\end{equation}
To compute the first term in Eq.~\eqref{eq:effective-second-order-gait}, we use the general radial squirming solution of Pak and Lauga
\cite{pak_generalized_2014}. 

\subsection{Radial derivative of the non-axisymmetric first-order flow}
\label{sec:radial-derivative-pak-lauga}

To pull the boundary condition from the deformed surface onto the reference
sphere, we require the radial derivative of the first-order flow at $r=a$.
Our deformation is non-axisymmetric, so we use the general solution of
Pak and Lauga for arbitrary radial surface velocities
\cite{pak_generalized_2014}.

We consider first a single spherical-harmonic mode of degree $\ell\geq2$
and azimuthal index $0\leq m\leq\ell$:
\begin{equation}
    \mathcal Y_{\ell m}(\vartheta,\varphi)
    =
    P_\ell^m(\cos\vartheta)e^{im\varphi},
    \label{eq:unnormalized-harmonic}
\end{equation}
where $P_\ell^m$ is a Legendre function. As is our setting here Eq.~\eqref{eq:effective-second-order-gait}, we prescribe purely
radial motion on the undeformed sphere,
\begin{equation}
    \mathbf u(a,\vartheta,\varphi)
    =
    V_{\ell m}\mathcal Y_{\ell m}(\vartheta,\varphi)\,
    \mathbf e_r,
    \label{eq:nonaxisymmetric-radial-bc}
\end{equation}
with vanishing polar and azimuthal surface velocities.

Pak and Lauga formulate the general surface motion in a real basis,
\begin{align}
    u_r(a,\vartheta,\varphi)
    &=
    \sum_{m\geq0}
    \left[
        D_m(\vartheta)\cos(m\varphi)
        +
        \widetilde D_m(\vartheta)\sin(m\varphi)
    \right],
    \\
    u_\vartheta(a,\vartheta,\varphi)
    &=
    \sum_{m\geq0}
    \left[
        E_m(\vartheta)\cos(m\varphi)
        +
        \widetilde E_m(\vartheta)\sin(m\varphi)
    \right],
    \\
    u_\varphi(a,\vartheta,\varphi)
    &=
    \sum_{m\geq0}
    \left[
        F_m(\vartheta)\cos(m\varphi)
        +
        \widetilde F_m(\vartheta)\sin(m\varphi)
    \right].
\end{align}
For the cosine component of
\eqref{eq:nonaxisymmetric-radial-bc}, one sets
\begin{equation}
    D_m(\vartheta)
    =
    V_{\ell m}P_\ell^m(\cos\vartheta),
    \qquad
    E_m=F_m=0,
\end{equation}
with all other modes equal to zero. The sine component is obtained
analogously using $\widetilde D_m$.

Equations~(165), (167), and (169) of
Ref.~\cite{pak_generalized_2014} then give the corresponding
coefficients in Lamb's general solution:
\begin{align}
    A_{m\ell}
    &=
    \frac{(2\ell-1)a^\ell\eta_{\rm f}}{2}
    \frac{(2\ell+1)(\ell-m)!}
    {(\ell+1)(\ell+m)!}
    \int_{-1}^{1}
    \ell V_{\ell m}
    \left[P_\ell^m(\mu)\right]^2\,{\rm d}\mu,
    \label{eq:Amell-integral}
    \\
    B_{m\ell}
    &=
    \frac{a^{\ell+2}}{4}
    \frac{(2\ell+1)(\ell-m)!}
    {(\ell+1)(\ell+m)!}
    \int_{-1}^{1}
    (\ell-2)V_{\ell m}
    \left[P_\ell^m(\mu)\right]^2\,{\rm d}\mu,
    \label{eq:Bmell-integral}
    \\
    C_{m\ell}&=0,
\end{align}
where $\eta_{\rm f}$ denotes the water viscosity. Using the
relation
\begin{equation}
    \int_{-1}^{1}
    \left[P_\ell^m(\mu)\right]^2\,{\rm d}\mu
    =
    \frac{2}{2\ell+1}
    \frac{(\ell+m)!}{(\ell-m)!},
    \label{eq:associated-legendre-norm}
\end{equation}
we obtain
\begin{equation}
    A_{m\ell}
    =
    a^\ell\eta_{\rm f}
    \frac{\ell(2\ell-1)}{\ell+1}
    V_{\ell m},
    \qquad
    B_{m\ell}
    =
    a^{\ell+2}
    \frac{\ell-2}{2(\ell+1)}
    V_{\ell m},
    \qquad
    C_{m\ell}=0.
    \label{eq:pure-radial-lamb-coefficients}
\end{equation}
Surprisingly, the dependence on $m$ cancels exactly between the projection prefactors and
the norm of the associated Legendre function.

Substituting \eqref{eq:pure-radial-lamb-coefficients} into the general
non-axisymmetric flow, Eqs.~(158)--(160) of
Ref.~\cite{pak_generalized_2014} and using the fact that $\ell \geq 2$ gives
\begin{equation}
    u_r(r,\vartheta,\varphi)
    =
    \frac{V_{\ell m}}{2}
    \left[
        \ell\left(\frac{a}{r}\right)^\ell
        -
        (\ell-2)\left(\frac{a}{r}\right)^{\ell+2}
    \right]
    \mathcal Y_{\ell m}(\vartheta,\varphi),
    \label{eq:nonaxisymmetric-radial-flow}
\end{equation}
while the tangential part is
\begin{equation}
    \mathbf u_\parallel(r,\vartheta,\varphi)
    =
    -
    \frac{\ell-2}{2(\ell+1)}
    V_{\ell m}
    \left[
        \left(\frac{a}{r}\right)^\ell
        -
        \left(\frac{a}{r}\right)^{\ell+2}
    \right]
    \nabla_\Omega\mathcal Y_{\ell m}(\vartheta,\varphi),
    \label{eq:nonaxisymmetric-tangential-flow}
\end{equation}
where we recall that
\begin{equation}
    \nabla_\Omega
    =
    \mathbf e_\vartheta\partial_\vartheta
    +
    \frac{\mathbf e_\varphi}{\sin\vartheta}\partial_\varphi
\end{equation}
is the angular gradient on the unit sphere. As a check, equations
\eqref{eq:nonaxisymmetric-radial-flow} and
\eqref{eq:nonaxisymmetric-tangential-flow} satisfy
\begin{equation}
    u_r(a)=V_{\ell m}\mathcal Y_{\ell m},
    \qquad
    \mathbf u_\parallel(a)=\mathbf 0,
\end{equation}
as required.

Differentiating the radial component at the sphere surface gives
\begin{align}
    a\left.\partial_r u_r\right|_{r=a}
    &=
    \frac{V_{\ell m}}{2}
    \left[
        -\ell^2+(\ell-2)(\ell+2)
    \right]\mathcal Y_{\ell m}
    \nonumber\\
    &=
    -2V_{\ell m}\mathcal Y_{\ell m}.
    \label{eq:nonaxisymmetric-radial-derivative-r}
\end{align}
Similarly,
\begin{align}
    a\left.\partial_r\mathbf u_\parallel\right|_{r=a}
    &=
    -
    \frac{\ell-2}{2(\ell+1)}
    V_{\ell m}
    \left[-\ell+(\ell+2)\right]
    \nabla_\Omega\mathcal Y_{\ell m}
    \nonumber\\
    &=
    -
    \frac{\ell-2}{\ell+1}
    V_{\ell m}\nabla_\Omega\mathcal Y_{\ell m}.
    \label{eq:nonaxisymmetric-radial-derivative-tangent}
\end{align}
We therefore obtain
\begin{equation}
    \boxed{
    a\left.\partial_r\mathbf u_{\ell m}\right|_{r=a}
    =
    -2V_{\ell m}\mathcal Y_{\ell m}\mathbf e_r
    -
    \frac{\ell-2}{\ell+1}
    V_{\ell m}\nabla_\Omega\mathcal Y_{\ell m}.
    }
    \label{eq:nonaxisymmetric-radial-derivative}
\end{equation}

We now use Eq.~\eqref{eq:nonaxisymmetric-radial-derivative} for our case in hand
\begin{equation}
    F_n(\vartheta,\varphi)
    =
    \sin^n\vartheta\,e^{in\varphi}.
\end{equation}
Since $F_n$ is proportional to $\mathcal{Y}_{n,n}$, 
% $\mathcal Y_{nn}$.
% \begin{equation}
%     P_n^n(\cos\vartheta)
%     =
%     (-1)^n(2n-1)!!
%     \sin^n\vartheta,
% \end{equation}
% $F_n$ is proportional to the sectoral spherical harmonic
% $\mathcal Y_{nn}$. 
Equation
\eqref{eq:nonaxisymmetric-radial-derivative} therefore applies with
$\ell=m=n$ by linearity. For the first-order boundary velocity
\begin{equation}
    \mathbf u^{(1)}(a)
    =
    a\dot f\,\mathbf e_r,
    \qquad
    f
    =
    \sum_{n\geq2}
    \left[
        \alpha_nF_n
        +
        \overline{\alpha_n}\,\overline{F_n}
    \right],
\end{equation}
we obtain mode by mode
\begin{equation}
    a\left.\partial_r\mathbf u_n^{(1)}\right|_{r=a}
    =
    -2a\dot\alpha_nF_n\,\mathbf e_r
    -
    a\frac{n-2}{n+1}
    \dot\alpha_n\nabla_\Omega F_n,
    \label{eq:sectoral-radial-derivative}
\end{equation}
together with its complex conjugate.

For briefness, we introduce the diagonal (infinite) matrix
\begin{equation}
    \mathcal QF_n
    =
    \frac{n-2}{n+1}F_n,
    \qquad
    \mathcal Q\overline{F_n}
    =
    \frac{n-2}{n+1}\overline{F_n},
    \label{eq:sectoral-Q-operator}
\end{equation}
so we can write
\begin{equation}
    \boxed{
    \left.\partial_r\mathbf u^{(1)}\right|_{r=a}
    =
    -2\dot f\,\mathbf e_r
    -
    \nabla_\Omega\mathcal Q\dot f.
    }
    \label{eq:radial-derivative-compact}
\end{equation}

\subsection{Determination of the swimming velocity}
We now return to the computation of $\mathbf{U}$. 

Substitution into \eqref{eq:effective-second-order-gait} yields
\begin{equation}
    \mathbf v_{\rm eff}^{(2)}
    =
    2af\dot f\,\mathbf e_r
    +
    af\nabla_\Omega\mathcal Q\dot f
    -
    a\dot f\,\nabla_\Omega f.
    \label{eq:effective-gait-expanded}
\end{equation}

For a sphere, the reciprocal theorem reduces the swimming velocity to the
surface average of the prescribed velocity (thanks to the stress $\sigma$ being constant for a rigid translating sphere)
\cite{stonePropulsionMicroorganisms}
\begin{equation}
    \mathbf U^{(2)}
    =
    -\frac{1}{4\pi a^2}
    \int_{S_0}\mathbf v_{\rm eff}^{(2)}\,{\rm d}S
    =
    -\frac{1}{4\pi}
    \int_{\mathbb S^2}\mathbf v_{\rm eff}^{(2)}\,{\rm d}\Omega.
    \label{eq:sphere-reciprocal}
\end{equation}
Using the identity (which can be proven simply by computing the gradient of $r g(\vartheta,\phi)$ and integrating over the unit ball)
\begin{equation}
    \int_{\mathbb S^2}\nabla_\Omega g\,{\rm d}\Omega
    =
    2\int_{\mathbb S^2}g\,\mathbf e_r\,{\rm d}\Omega,
    \label{eq:spherical-gradient-identity}
\end{equation}
one finds
\begin{equation}
    \int_{\mathbb S^2}
    \left(
        2f\dot f\,\mathbf e_r
        -
        \dot f\,\nabla_\Omega f
    \right)
    {\rm d}\Omega
    =
    \int_{\mathbb S^2}
    f\nabla_\Omega\dot f\,{\rm d}\Omega.
\end{equation}
Hence,
\begin{equation}
    \mathbf U^{(2)}
    =
    -\frac{a}{4\pi}
    \int_{\mathbb S^2}
    f\nabla_\Omega\mathcal (1+\mathcal Q)\dot f\,{\rm d}\Omega,
    \label{eq:compact-sphere-velocity}
\end{equation}
where again, 
\begin{equation}
    \mathcal (1+\mathcal Q)F_n
    =
    \frac{2n-1}{n+1} F_n.
    \label{eq:D-eigenvalue}
\end{equation}

It now remains to evaluate the integral \eqref{eq:compact-sphere-velocity}. 
We use the complex representation
\begin{equation}
    \mathcal U\equiv U_x+iU_y.
\end{equation}
We have for our choice $F_n = \sin^n (\theta) e^{in\varphi}$: % YES Checked, both equalities
\begin{align}
    \left(\nabla_\Omega F_n\right)_x
    +i\left(\nabla_\Omega F_n\right)_y
    &=
    -nF_{n+1},
    \label{eq:grad-sectoral-plus}\\
    \left(\nabla_\Omega\overline{F_{n+1}}\right)_x
    +i\left(\nabla_\Omega\overline{F_{n+1}}\right)_y
    &=
    (n+1)\sin^n\vartheta
    \left(1+\cos^2\vartheta\right)e^{-in\varphi} = (n+1) (1+\cos^2 \vartheta) \overline{F_n}.
    \label{eq:grad-sectoral-minus}
\end{align}
Thus 
\begin{align}
    f\nabla_\Omega\mathcal (1+\mathcal Q)\dot f &= \left(\sum_{n \geq 2} \left[\alpha_n F_n + \overline{\alpha_n F_n} \right]\right)\nabla_\Omega\left(\sum_{m \geq 2} \frac{2m-1}{m+1} \left[\dot \alpha_m F_m + \overline{\dot \alpha_m F_m} \right]\right) \\ &=\left(\sum_{n \geq 2} \left[\alpha_n F_n + \overline{\alpha_n F_n} \right]\right)\left(\sum_{m \geq 2} \frac{2m-1}{m+1}\left[-m \dot \alpha_m F_{m+1} + m(1+\cos^2 \vartheta) \overline{\dot \alpha_m F_{m-1}} \right]\right)
\end{align}

By orthogonality of the spherical harmonics $F_n$ over the angle $\phi$, inserting the above in the integral in \eqref{eq:compact-sphere-velocity} yields 
vanishing contributions unless the terms $F_n$ and $\overline{F_{m - 1}}$ are coupled by $n=m - 1$, and $\overline{F_n}$ and $F_{m+1}$ are coupled by $n=m+1$.

Defining
\begin{equation}
    I_n
    \equiv
    \int_0^\pi\sin^{2n+1}\vartheta\,{\rm d}\vartheta
    =
    \frac{2^{2n+1}(n!)^2}{(2n+1)!}, \quad \int_0^\pi(1+\cos^2 \vartheta)\sin^{2n+1}\vartheta\,{\rm d}\vartheta = 2I_n -I_{n+1},
    \label{eq:In}
\end{equation}
we obtain, after some algebra and after restoring the prefactor $\varepsilon^2$, % YES Checked.
\begin{equation}
    \boxed{
    \mathcal U_{\rm geom}
    =
    -a\varepsilon^2
    \sum_{n\geq2}
    \left[
        A_n\alpha_n\overline{\dot\alpha_{n+1}}
        +
        B_n\overline{\alpha_{n+1}}\dot\alpha_n
    \right]
    +O(\varepsilon^3),
    }
    \label{eq:spherical-instantaneous-velocity}
\end{equation}
with
\begin{align}
    A_n
    &=
    \frac{(n+1)(2n+1)}{2n+3}\,I_n
    =
    \frac{
        2^{2n+1}(n!)^2(n+1)(2n+1)
    }{
        (2n+3)(2n+1)!
    },
    \label{eq:An-sphere}\\
    B_n
    &=
    -\frac{n(2n-1)}{2n+3}\,I_n
    =
    -\frac{
        2^{2n+1}(n!)^2n(2n-1)
    }{
        (2n+3)(2n+1)!
    }.
    \label{eq:Bn-sphere}
\end{align}

For a periodic stroke, integration by parts over one period gives
\begin{equation}
    \left\langle
        \overline{\alpha_{n+1}}\dot\alpha_n
    \right\rangle
    =
    -
    \left\langle
        \alpha_n\overline{\dot\alpha_{n+1}}
    \right\rangle.
\end{equation}
The cycle-averaged velocity consequently reduces to
\begin{equation}
    \boxed{
    \left\langle\mathcal U_{\rm geom}\right\rangle
    =
    -a\varepsilon^2
    \sum_{n\geq2}
    C_n
    \left\langle
        \alpha_n\overline{\dot\alpha_{n+1}}
    \right\rangle,
    }
    \label{eq:spherical-mean-velocity}
\end{equation}
where
\begin{equation}
    C_n
    =
    A_n-B_n
    =
    \frac{
        2^{2n+1}(n!)^2
        \left(4n^2+2n+1\right)
    }{
        (2n+3)(2n+1)!
    }.
    \label{eq:Cn-sphere}
\end{equation}

\subsection{An explicit choice for the deformation}
\begin{figure}
    \centering
    \includegraphics[width=.24\textwidth]{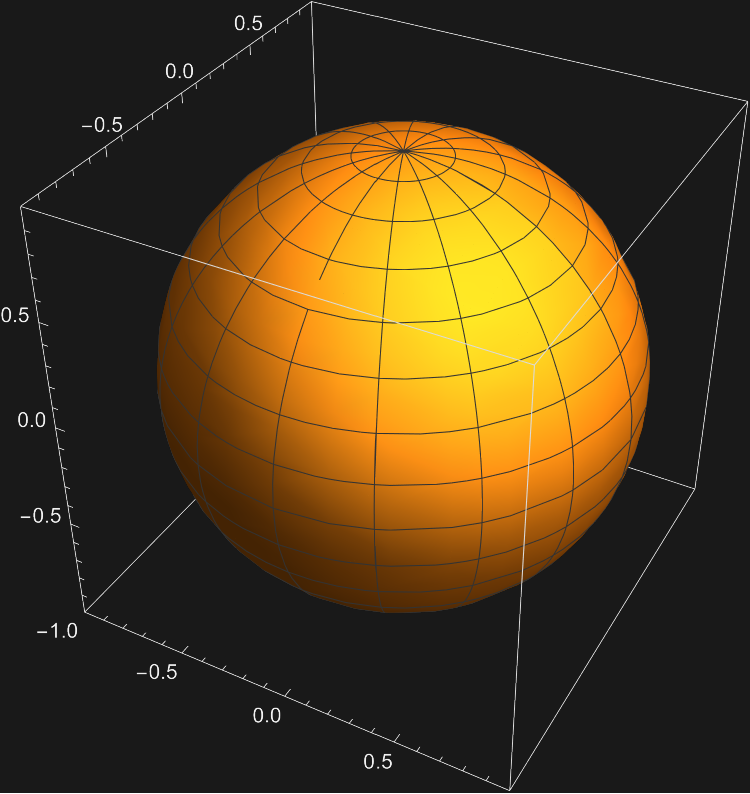}
    \includegraphics[width=.24\textwidth]{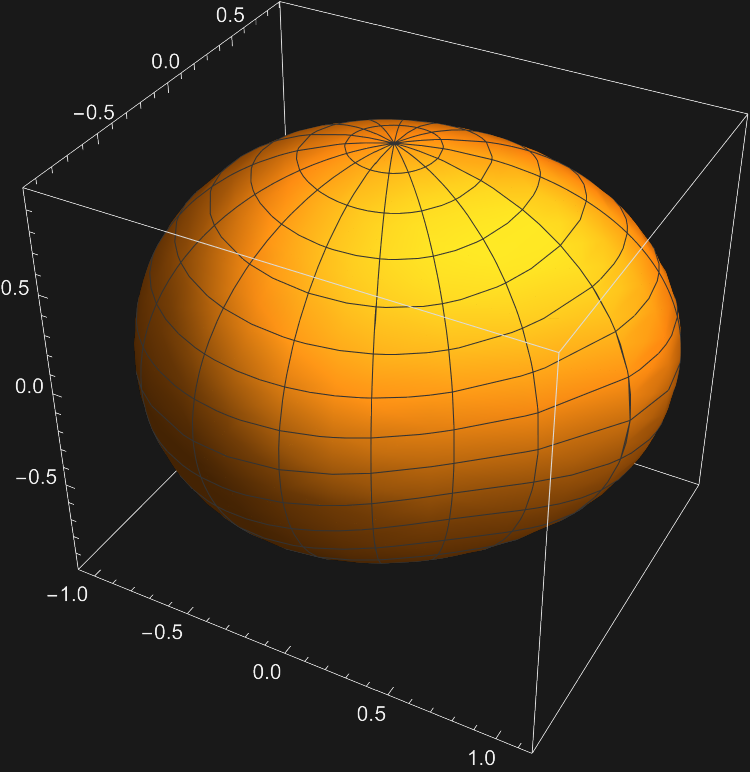}
    \includegraphics[width=.24\textwidth]{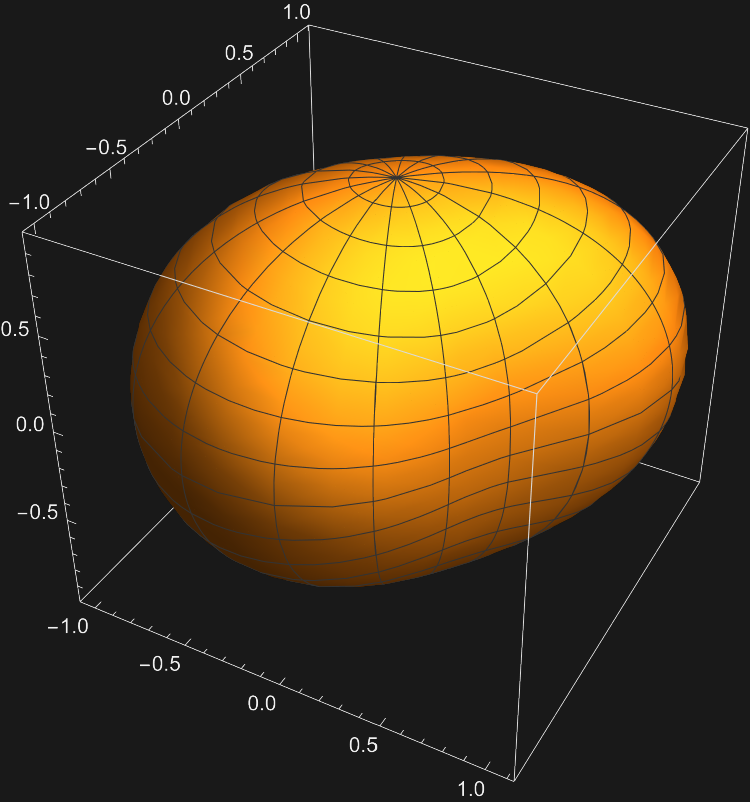}
    \includegraphics[width=.24\textwidth]{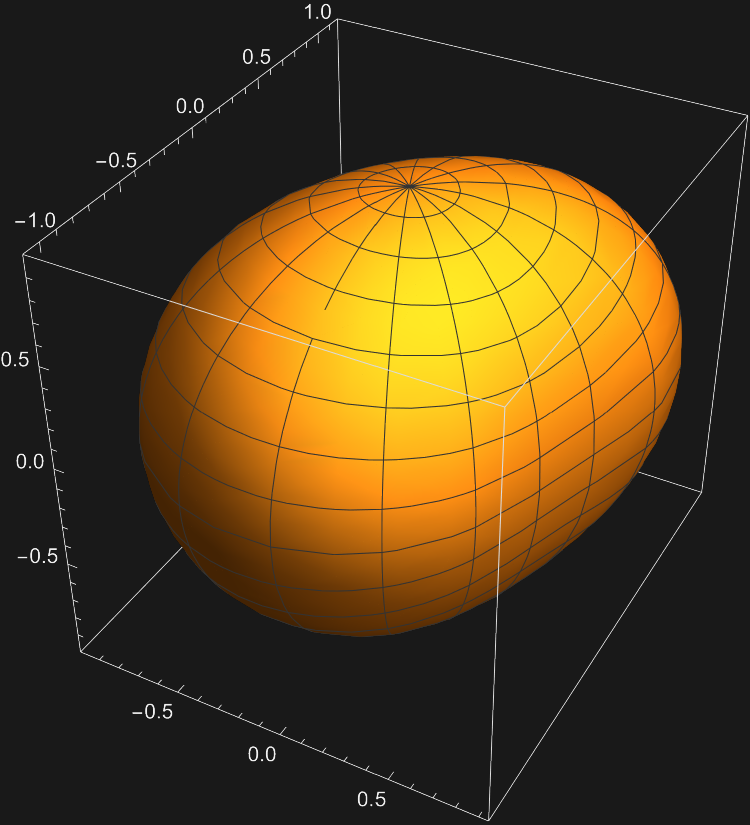}
    \caption{Our choice of deformation in the spherical droplet approximation. The figures show the deformation from $t=0$ to $t=T/4$, with $T$ being the period of the stroke; from left to right, $t=0, \frac{T}{12}, \frac{T}{6}, \frac{3T}{12}$. Then, from $T/4$ to $T$, the sphere is not deformed. This mimics the traveling bump produced by the raft. Note that our results in the main text, in particular our comparisons with experimental data, are independent of our choice of deformation here, which serves to illustrate our computations above.  }
    \label{fig:deform-example}
\end{figure}
In this section, we propose an explicit deformation pattern in the spherical droplet approximation to illustrate the swimming mechanism. Choosing a stroke period $T$ and a real $0<\alpha<1$, we choose the deformation
\begin{equation}
    f(\theta, \varphi, t) = \begin{cases}\sum_{n=2}^\infty \frac{\sin^n \theta}{n!} \cos(n(\varphi- \omega t)), \quad &t \mod T < \alpha T, \\ 0, \quad &t \mod T >\alpha T.
    \end{cases}
\end{equation}
(in practice, we need a smoothed-out version of the above, to avoid introducing discontinuities. This does not change the discussion.) 
In the notations of the section above, 
\begin{equation}
    F_n(\theta, \varphi) = \sin^n \theta e^{i n \varphi}, \quad \alpha_n(t) = \frac{1}{2} \frac{e^{-i n\omega t}}{n!}, \quad \omega = \frac{2\pi}{T}. 
\end{equation}
It can be easily shown that 
\begin{equation}
    f(\theta, \varphi,t) = \begin{cases}e^{\sin (\theta ) \cos (\varphi -\omega t)} \cos (\sin (\theta ) \sin (\varphi -\omega t))-\sin (\theta ) \cos (\varphi - \omega t)-1, \quad &t \mod T < \alpha T, \\ 0, \quad &t \mod T >\alpha T.
    \end{cases}
\end{equation}
This deformation is plotted Fig.~\ref{fig:deform-example}. It possesses the main features relevant to our experimental setting: 
\begin{enumerate}
    \item It is nonzero only during a certain time window, here $0<t \mod T < \alpha T$. In other words, a fraction $1-\alpha$ of the time, the sphere is undeformed (choose e.g. $\alpha = \frac{1}{4}$ for definiteness).
    \item It is strongest at the equator $\theta \approx \frac{\pi}{2}$. 
    \item The associated bump travels along the equator at frequency $\omega$ (the raft's angular velocity).
    \item Finally (this is a technical choice for simplicity and compatibility with our derivation above), because we choose to expand on the diagonal spherical harmonic basis $F_n$, our result Eq.~\eqref{eq:spherical-mean-velocity} is directly applicable to compute the swimming velocity.
\end{enumerate}
For this deformation pattern, we find the swimming velocity 
\begin{equation}
    \mathbf{U}_{\rm geom} \approx -0.2109087014 \times a \varepsilon^2 \times i \omega e^{i \omega t}.
\end{equation}
Over a period $T$, one then has the average swimming velocity 
\begin{equation}
    \langle \mathbf{U}_{\rm geom} \rangle \approx -0.21 \times a \varepsilon^2 \times \frac{\int_0^{\alpha T \omega} i e^{i u} du}{T} = -0.21\times a \varepsilon^2 \times \frac{e^{2i \pi \alpha}-1}{T}.
\end{equation}
For $\alpha = \frac{1}{4}$, one has 
\begin{equation}
    \langle \mathbf{U}_{\rm geom}\left(\alpha =\tfrac{1}{4}\right) \rangle \approx 0.21 \times a \varepsilon^2 \times \frac{1-i}{T}.
\end{equation}
The droplet is thus propelled in the direction $\mathbf{e_x}-\mathbf{e_y}$.

\section{Mean-field model for the internal colloid cluster dynamics}
%==============================================================================

We now construct a minimal dynamical model for the active colloid cluster confined inside the droplet. The cluster's motion will be responsible for the deformation $\xi(\theta,t)$ of the droplet interface that was prescribed phenomenologically in the above section. The model is built from physical assumptions at the scale of the individual colloids in the same spirit as \cite{Kato2025}.

%------------------------------------------------------------------------------
\subsection{Geometry and degrees of freedom}
%------------------------------------------------------------------------------

The cluster is a rigid assembly of $N$ identical colloids. Its centre of mass is
\begin{equation}
  \mathbf{R}(t) = R(t)\,\mathbf{e}_r\!\bigl(\phi(t)\bigr),
\end{equation}
where $\phi$ is the polar angle measured from the laboratory frame axis $\mathbf{e}_x$. The orientation of the rigid body (a chosen axis along the solid) is given by the absolute angle $\beta(t)$ with respect to the horizontal axis $\mathbf{e}_x$.  
Colloid $i$ has a fixed position in the body frame,
\begin{equation}
  \mathbf{r}_i = \mathbf{R} + d_i\,\mathbf{e}_r(\beta + \alpha_i),
\end{equation}
with constants $d_i\ge 0$, $\alpha_i$ satisfying $\sum_i d_i e^{i\alpha_i}=0$ by definition of the centre of mass. Finally, due to the asymmetric coating, each colloid carries a polarity vector
\begin{equation}
  \mathbf{p}_i = \mathbf{e}_r(\beta + \theta_i),
\end{equation}
where $\theta_i(t)$ is the orientation of its active driving force relative to the body axis.

The collective polarization is thus
\begin{equation}
  \mathbf{P} = \frac{1}{N}\sum_{i=1}^N \mathbf{p}_i .
\end{equation}

%------------------------------------------------------------------------------
%------------------------------------------------------------------------------
\subsection{Forces and torques on a single colloid}
%------------------------------------------------------------------------------

Every colloid $i$ is subject to four distinct actions:

\begin{enumerate}
  \item \textbf{Active self-propulsion.}
        A colloid exerts a force
        $\mathbf{F}_i^{\rm act,fluid}=-\xi v_p\mathbf{p}_i$
        on the surrounding fluid and experiences the opposite effective
        reaction,
        \begin{equation}
        \mathbf{F}_i^{\rm act}
        =
        \xi v_p\mathbf{p}_i.
        \end{equation}
        Here, $v_p\mathbf{p}_i$ is the active velocity of the
        isolated colloid. 

  \item \textbf{Confinement of the flow.}
        Let $\mathbf{u}^{\infty}_i$ denote the flow seen by
        colloid $i$, excluding the disturbance produced by the colloid
        itself. Fax\'en's law gives
        \begin{equation}
        \mathbf{F}_i^{\rm drag}
        =
        -\xi
        \left[
        \dot{\mathbf r}_i
        -
        \mathbf{u}^{\infty}_i(\mathbf r_i)
        -
        \frac{a_p^2}{6}
        \nabla^2\mathbf{u}^{\infty}_i(\mathbf r_i)
        \right],
        \label{eq:faxen-force}
        \end{equation}
        where $a_p$ is the colloid radius.

        Because the raft is a solid, the velocity of the center of colloid $i$ is
        \begin{equation}
        \dot{\mathbf r}_i
        =
        \dot{\mathbf R}
        +
        \dot\beta\,\mathbf e_z\times\mathbf q_i,
        \qquad
        \mathbf q_i=\mathbf r_i-\mathbf R.
        \end{equation}
        Rotation of the raft induces flow in
        the surrounding liquid (whether it is the oil inside the droplet, or the water below it). Because the droplet is finite, reflection of
        this disturbance by its boundary produces a reaction
        flow. Expanding this flow around the raft center reads at leading order
        \begin{equation}
        \mathbf{u}^{\infty}_i(\mathbf R+\mathbf q_i)
        -
        \mathbf{u}^{\infty}_i(\mathbf R)
        =
        c\dot\beta\,
        \mathbf e_z\times\mathbf q_i
        +
        O\left(\frac{q_i^2}{L^2}\right),
        \label{eq:local-reaction-flow}
        \end{equation}
        where $L$ is the droplet size and $c=c(L)$ is a
        confinement-dependent coefficient measuring the strength of the flow due to the raft's rotation that is reflected by the droplet's boundary. Most importantly, $c$ vanishes in the large-$L$ limit. The common
        translational velocity
        $\mathbf{u}^{\infty}_i(\mathbf R)$ is finally absorbed into the effective
        center-of-mass mobility $\xi$ introduced below.

        The field kept in Eq.~\eqref{eq:local-reaction-flow} is affine in
        $\mathbf q_i$, and therefore
        \begin{equation}
        \nabla^2\mathbf{u}^{\infty}_i=0.
        \end{equation}
        Eq.~\eqref{eq:faxen-force} thus leads to
        \begin{equation}
        \boxed{
        \mathbf{F}_i^{\rm drag}
        =
        -\xi
        \left[
        \dot{\mathbf R}
        +
        (1-c)\dot\beta\,
        \mathbf e_z\times\mathbf q_i
        \right].
        }
        \label{eq:particle-drag-faxen}
        \end{equation}
        % Thus the rotational reaction flow affects the drag on each colloid,
        % although its contribution to the total force cancels when summed over
        % the raft because $\sum_i\mathbf q_i=\mathbf 0$.

  \item \textbf{Isotropic confinement.}
        The droplet interface provides a capillary restoring force. We
        describe it by a generic isotropic single-particle potential $U(r)$,
        with a unique minimum at the droplet center:
        \begin{equation}
        \mathbf{F}_i^{\rm conf}
        =
        -\boldsymbol\nabla_{\mathbf r_i}U(r_i)
        =
        -\frac{U'(r_i)}{r_i}\mathbf r_i,
        \qquad
        r_i=|\mathbf r_i|.
        \label{eq:general-confining-force}
        \end{equation}
        The
        harmonic potential used in the main text,
        \begin{equation}
        U(r)=\frac{k}{2}r^2,
        \end{equation}
        is a simple representative and gives
        $\mathbf F_i^{\rm conf}=-k\mathbf r_i$~\cite{Kato2025}.
        Below, we keep $U$ general in order to distinguish the consequences
        of isotropic confinement from those that are specific to harmonic
        confinement.

  \item \textbf{Marangoni (capillary) torque.}
        Each colloid is partially wetted at the interface and is propelled by
        a Marangoni flow generated by the temperature gradients resulting
        from its asymmetric heating. The local gradient is predominantly
        oriented along the collective polarization $\mathbf P$. The torque
        $\tau_i$ on colloid $i$ must be odd under
        $\mathbf P\to-\mathbf P$, increase with the distance $r_i$ from the
        droplet center~\cite{Kato2025}, and vanish when the local gradient is
        radial, i.e. when $\mathbf P\parallel\mathbf r_i$, since the colloid
        should not rotate in this configuration. The simplest admissible form
        is therefore
        \begin{equation}
        \tau_i
        =
        \Gamma
        \bigl(\mathbf r_i\times\mathbf P\bigr)
        \cdot\mathbf e_z,
        \label{eq:marangoni-torque-single}
        \end{equation}
        where $\Gamma$ is a phenomenological coefficient. Higher-order terms
        may in principle be present, but we deliberately restrict ourselves
        to this leading rotationally invariant term.
\end{enumerate}

\subsubsection{Relative kinetics of the fluid and the solid raft}

The raft is rigid with respect to the relative positions of the colloid
centers, but individual colloids retain an internal rotational degree of
freedom. We therefore distinguish the rigid-body rotation $\beta$ of the
positional raft from the rotation $\theta_i$ of colloid $i$ relative to it.
The absolute orientation of its polarity is $\beta+\theta_i$.

The rotational Fax\'en law states that the viscous torque depends on the
angular velocity of the colloid relative to the regular ambient vorticity:
\begin{equation}
\tau_i^{\rm drag}
=
-\xi_r
\left[
\dot\beta+\dot \theta_i
-
\frac{1}{2}
\bigl(\boldsymbol\nabla\times
\mathbf u_i^\infty\bigr)_{\mathbf r_i}
\cdot\mathbf e_z
\right],
\label{eq:rotational-faxen}
\end{equation}
where $\xi_r$ is the effective angular resistance of one colloid.

The vorticity of the reaction flow in
Eq.~\eqref{eq:local-reaction-flow} writes
\begin{equation}
\frac{1}{2}
\boldsymbol\nabla\times\mathbf u_i^\infty
=
c\dot\beta\,\mathbf e_z.
\end{equation}
Equation~\eqref{eq:rotational-faxen} consequently gives
\begin{equation}
\boxed{
\tau_i^{\rm drag}
=
-\xi_r
\left[
\dot\theta_i+(1-c)\dot\beta
\right].
}
\label{eq:rotational-drag-final}
\end{equation}
The same coefficient $c$ therefore reduces both the rotational slip of the
colloid centers and the kinetics of each individual colloid's polarity relative to the
ambient liquid. 

\vspace{1.5ex}
Let us make a subtle but important remark here. Motion of the raft center (via its orbital angle $\phi$) may also generate vorticity in the
droplet. The leading rotationally invariant
contribution is proportional to
$(\mathbf R\times\dot{\mathbf R})\cdot\mathbf e_z$. For harmonic confinement,
the center-of-mass equation derived below gives
\begin{equation}
(\mathbf R\times\dot{\mathbf R})\cdot\mathbf e_z
=
v_p
(\mathbf R\times\mathbf P)\cdot\mathbf e_z,
\end{equation}
so that such a contribution has the same form as the collective Marangoni
torque and can be absorbed into the phenomenological coefficient $\Gamma$.
For a general particle-wise potential $U$, however, the total confining force
need not be exactly parallel to $\mathbf R$, and this identification is no
longer exact. For clarity, and because this is not used in the main text, we do not introduce a separate orbital term
in the minimal model. The coefficient $\Gamma$ should then be understood as the
effective strength of the leading torque retained in
Eq.~\eqref{eq:marangoni-torque-single}.

%------------------------------------------------------------------------------
\subsection{Dynamics of the centre of mass}
%------------------------------------------------------------------------------

The overdamped force balance on the whole cluster reads
\begin{equation}
\sum_{i=1}^N
\left[
\xi v_p\mathbf p_i
-
\xi
\left(
\dot{\mathbf R}
+
(1-c)\dot\beta\,
\mathbf e_z\times\mathbf q_i
\right)
-
\boldsymbol\nabla_{\mathbf r_i}U(r_i)
\right]
=
\mathbf 0.
\label{eq:full-force-balance}
\end{equation}
Since
\begin{equation}
\sum_i\mathbf q_i=\mathbf 0,
\end{equation}
the rotational reaction-flow contribution cancels from the total force.
Introducing the total confining energy
\begin{equation}
\mathcal U(\mathbf R,\beta)
=
\sum_{i=1}^N
U\bigl(|\mathbf R+\mathbf q_i|\bigr),
\label{eq:total-confining-energy}
\end{equation}
we have
\begin{equation}
\boldsymbol\nabla_{\mathbf R}\mathcal U
=
\sum_i
\boldsymbol\nabla_{\mathbf r_i}U(r_i).
\end{equation}
The center-of-mass dynamics is therefore
\begin{equation}
\boxed{
\dot{\mathbf R}
=
v_p\mathbf P
-
\frac{1}{N\xi}
\boldsymbol\nabla_{\mathbf R}\mathcal U(\mathbf R,\beta).
}
\label{eq:general-com-dynamics}
\end{equation}

Although the single-particle potential $U(r)$ is isotropic, the total energy
$\mathcal U(\mathbf R,\beta)$ generally depends on the raft orientation
$\beta$. For an extended asymmetric raft, its gradient need not be exactly
parallel to $\mathbf R$. In polar coordinates $(R,\phi)$, the general
center-of-mass equations are consequently
\begin{align}
\dot R
&=
v_p|\mathbf P|\cos\Psi
-
\frac{1}{N\xi}
\mathbf e_r\cdot
\boldsymbol\nabla_{\mathbf R}\mathcal U,
\label{eq:general-radial-dynamics}
\\
\dot\phi
&=
\frac{v_p|\mathbf P|\sin\Psi}{R}
-
\frac{1}{N\xi R}
\mathbf e_\phi\cdot
\boldsymbol\nabla_{\mathbf R}\mathcal U,
\label{eq:general-angular-dynamics}
\end{align}
where $\Psi(t)$ is the instantaneous angle between $\mathbf P$ and
$\mathbf e_r$.

Harmonic confinement is exceptional. For
\begin{equation}
U(r)=\frac{k}{2}r^2,
\end{equation}
the center-of-mass condition $\sum_i\mathbf q_i=\mathbf 0$ gives
\begin{equation}
\mathcal U(\mathbf R,\beta)
=
\frac{Nk}{2}R^2
+
\frac{k}{2}\sum_iq_i^2,
\end{equation}
which is independent of $\beta$. Hence
\begin{equation}
\boldsymbol\nabla_{\mathbf R}\mathcal U
=
Nk\mathbf R,
\end{equation}
and Eqs.~\eqref{eq:general-radial-dynamics}--%
\eqref{eq:general-angular-dynamics} reduce to the ones shown in the main text
\begin{equation}
\boxed{
\dot R
=
v_p|\mathbf P|\cos\Psi
-
\frac{k}{\xi}R,
\qquad
\dot\phi
=
\frac{v_p|\mathbf P|\sin\Psi}{R}.
}
\label{eq:harmonic-com-dynamics}
\end{equation}

%------------------------------------------------------------------------------
\subsection{Internal polarity dynamics and frequency synchronization}
%------------------------------------------------------------------------------

In the absence of an ambient reaction flow, the overdamped rotational
dynamics of a single colloid would obey
\begin{equation}
\xi_r(\dot\beta+\dot\theta_i)
=
\tau_i.
\label{torquei0}
\end{equation}
In a finite droplet, rotational Fax\'en's law instead gives the drag
Eq.~\eqref{eq:rotational-drag-final}. Including the individual Marangoni
torque and the dissipative coupling to neighboring colloids yields
\begin{equation}
\boxed{
\xi_r
\left[
\dot\theta_i+(1-c)\dot\beta
\right]
=
\Gamma
(\mathbf r_i\times\mathbf P)\cdot\mathbf e_z
+
\sum_{j\ {\rm neighbor\ to}\ i}
\tau'_{j\to i}.
}
\label{torquei}
\end{equation}

Viscous and contact interactions between neighboring colloids penalize
differences in angular velocities. We represent this by
frequency-equalizing torques, for example
\begin{equation}
\tau'_{i\to j}
=
-\kappa(\dot\theta_j-\dot\theta_i)
=
-\tau'_{j\to i}.
\end{equation}
These interactions tend to equalize the individual frequencies for large $\kappa$:
\begin{equation}
\dot\theta_i(t)=\dot\theta(t)
\quad\Longrightarrow\quad
\theta_i(t)=\theta(t)+\delta_i,
\end{equation}
where the constant phase lags $\delta_i$ retain the frozen initial
disorder of the polarities.

Summing Eq.~\eqref{torquei} over the colloids eliminates the total internal
torque because
\begin{equation}
\sum_i
\sum_{j\ {\rm neighbor\ to}\ i}
\tau'_{j\to i}
=
0.
\end{equation}
Moreover,
\begin{equation}
\sum_i
\mathbf r_i
=
N\mathbf R.
\end{equation}
The evolution of the common internal rotation is therefore
\begin{equation}
N\xi_r
\left[
\dot\theta+(1-c)\dot\beta
\right]
=
N\Gamma
(\mathbf R\times\mathbf P)\cdot\mathbf e_z,
\end{equation}
or
\begin{equation}
\boxed{
\dot\theta+(1-c)\dot\beta
=
\frac{\Gamma}{\xi_r}
(\mathbf R\times\mathbf P)\cdot\mathbf e_z.
}
\label{eq:locked-spin-dynamics}
\end{equation}

\textbf{Collective polarization.}
With the above synchronization hypothesis,
\begin{equation}
  \mathbf{P} = \frac{1}{N}\sum_i e^{i(\beta + \theta + \delta_i)}
            = e^{i(\beta + \theta)}\,\mathbf{P}_0,
\end{equation}
where 
\begin{equation}
  \mathbf{P}_0 = \frac{1}{N}\sum_i e^{i\delta_i} \equiv |\mathbf{P}_0|\,e^{i\delta}
\end{equation}
is a constant complex vector determined solely by the frozen disorder. Its magnitude $|\mathbf{P}_0| \le 1$ is a constant of motion. The angle between $\mathbf{P}$ and $\mathbf{e}_r$ is therefore
\begin{equation}
  \Psi \equiv \beta - \phi + \theta + \delta.
\end{equation}
We note here that minimal models where angular frequencies $\dot \theta_i(t)$ are not forced by internal torques to be equal to each other lead to a rapidly decaying polarization magnitude $|\mathbf{P}_0| \to 0$, which would mean that the cluster would stop moving (no local gradient of temperature at the interface). This is not seen in the experiments and motivates our synchronization ansatz.

%------------------------------------------------------------------------------
%------------------------------------------------------------------------------
\subsection{Rigid-body rotation of the cluster}
%------------------------------------------------------------------------------

To obtain the rigid-body rotation $\dot\beta$, we consider the total moment of
the forces acting on the particle centers about the raft center
$\mathbf R$. Internal constraint forces maintain the fixed relative positions
of the colloids. Their action--reaction pairs have zero total force and zero
total moment on the complete raft, and therefore do not enter the global
balance.

Another important point: the Marangoni torque $\tau_i$ introduced in
Eq.~\eqref{eq:marangoni-torque-single} acts about the center of colloid $i$ and
rotates its internal polarity $\beta+\theta_i$. Since the colloids
remain free to spin relative to the positional skeleton, this torque enters
the internal dynamics derived in the sections above but does not act directly on the
rigid-body coordinate $\beta$. 

The remaining contributions are the moments of the active propulsion forces,
the viscous forces on the particle centers, and the confining forces. Using
the force derived in Eq.~\eqref{eq:particle-drag-faxen}, the total
moment balance reads
\begin{align}
\mathbf 0
={}&
\sum_{i=1}^N
\mathbf q_i\times
\left[
\xi v_p\mathbf p_i
-
\xi
\left(
\dot{\mathbf R}
+
(1-c)\dot\beta\,
\mathbf e_z\times\mathbf q_i
\right)
-
\boldsymbol\nabla_{\mathbf r_i}U(r_i)
\right].
\label{eq:full-body-torque-balance}
\end{align}

The common translational velocity does not contribute to the moment because
\begin{equation}
\sum_i
\mathbf q_i\times\dot{\mathbf R}
=
\left(\sum_i\mathbf q_i\right)
\times\dot{\mathbf R}
=
\mathbf 0.
\end{equation}
For the rotational part of the drag,
\begin{equation}
\mathbf q_i\times
\left(
\mathbf e_z\times\mathbf q_i
\right)
=
q_i^2\mathbf e_z,
\end{equation}
so that
\begin{equation}
\sum_i
\mathbf q_i\times
\left[
-\xi(1-c)\dot\beta\,
\mathbf e_z\times\mathbf q_i
\right]
=
-\xi(1-c)I\dot\beta\,\mathbf e_z,
\end{equation}
where
\begin{equation}
I
=
\sum_{i=1}^N|\mathbf{q_i}|^2
=
\sum_{i=1}^Nd_i^2
\label{eq:geometrical-moment}
\end{equation}
is the geometrical rotational factor of the raft.

The confining contribution can be written directly in terms of the total
potential
\begin{equation}
\mathcal U(\mathbf R,\beta)
=
\sum_{i=1}^N
U\bigl(|\mathbf R+\mathbf q_i|\bigr),
\end{equation}
introduced in Eq.~\eqref{eq:total-confining-energy}. Since
\begin{equation}
\frac{\partial\mathbf q_i}{\partial\beta}
=
\mathbf e_z\times\mathbf q_i,
\end{equation}
one has
\begin{align}
\frac{\partial\mathcal U}{\partial\beta}
&=
\sum_i
\boldsymbol\nabla_{\mathbf r_i}U(r_i)
\cdot
\left(
\mathbf e_z\times\mathbf q_i
\right)
\nonumber\\
&=
\sum_i
\left[
\mathbf q_i\times
\boldsymbol\nabla_{\mathbf r_i}U(r_i)
\right]\cdot\mathbf e_z.
\end{align}
The moment of the confining forces is therefore
\begin{equation}
\boxed{
M_U
=
-\frac{\partial\mathcal U}{\partial\beta}
=
-\sum_i
\frac{U'(r_i)}{r_i}
\left(
(\mathbf{r_i-\mathbf{R}})\times\mathbf r_i
\right)\cdot\mathbf e_z = \left[\mathbf{R} \times \sum_i
\frac{U'(r_i)}{r_i}
\mathbf{q_i}\right]\cdot\mathbf e_z.
}
\label{eq:general-wall-torque}
\end{equation}
While the above vanishes for a harmonic potential as $U'(r_i) = k r_i$ and $\sum \mathbf{q_i} = \mathbf{0}$, it does not for a generic choice of $U$.

Collecting the different contributions in
Eq.~\eqref{eq:full-body-torque-balance} gives
\begin{equation}
\boxed{
\xi(1-c)I\dot\beta
=
\xi v_p
\sum_{i=1}^N
\left(
\mathbf q_i\times\mathbf p_i
\right)\cdot\mathbf e_z
-
\frac{\partial\mathcal U}{\partial\beta}.
}
\label{eq:general-body-dynamics}
\end{equation}

For the frequency-locked polarities,
\begin{equation}
\theta_i=\theta+\delta_i,
\end{equation}
we have
\begin{equation}
\left(
\mathbf q_i\times\mathbf p_i
\right)\cdot\mathbf e_z
=
d_i
\sin\left(
\theta+\delta_i-\alpha_i
\right).
\end{equation}
Introducing the complex geometrical moment
\begin{equation}
\boxed{
\sum_{i=1}^N
d_i
e^{i(\delta_i-\alpha_i)}
\equiv
\bar d\,e^{-i\bar\alpha},
\qquad
\bar d\geq0,
}
\label{eq:geometrical-active-moment}
\end{equation}
the total active moment becomes
\begin{equation}
\sum_i
\left(
\mathbf q_i\times\mathbf p_i
\right)\cdot\mathbf e_z
=
\bar d
\sin(\theta-\bar\alpha).
\end{equation}
Equation~\eqref{eq:general-body-dynamics} can therefore be written as
\begin{equation}
\boxed{
\dot\beta
=
\frac{
\xi v_p\bar d\sin(\theta-\bar\alpha)
-
\partial_\beta\mathcal U
}{
\xi(1-c)I
}.
}
\label{eq:body-dynamics-dimensional}
\end{equation}

%------------------------------------------------------------------------------
\subsection{Adimensionalized equations}
%------------------------------------------------------------------------------
In this section, we consider a purely harmonic potential $U(r_i) =(k/2) r_i^2$, as in the main text, so that $\partial_\beta \mathcal U = 0$. 
Collecting the equations for the three core variables $R,\Psi,\theta$ eliminates $\phi$ and $\beta$. Using \[\dot{\Psi} = \dot{\beta} - \dot{\phi} + \dot{\theta}, \quad \dot \beta = \frac{v_p \overline{d} \sin(\theta-\overline{\alpha})}{I(1-c)}\] we arrive at the closed dynamical system
\begin{equation}
\label{dyn-sys-dimens}
  \boxed{
    \begin{aligned}
      \dot{R} &= v_p|\mathbf{P}_0|\cos\Psi - \frac{k}{\xi}R,\\[4pt]
      \dot{\Psi} &= \frac{c v_p}{I(1-c)}\,\bar{d}\,\sin(\theta - \bar{\alpha})
                  + |\mathbf{P}_0|\Bigl[\frac{\Gamma}{\xi_r}R - \frac{v_p}{R}\Bigr]\sin\Psi,\\[4pt]
      \dot{\theta} &= \frac{\Gamma}{\xi_r}\,|\mathbf{P}_0|\,R\,\sin\Psi - \frac{v_p \overline{d} \sin(\theta-\overline{\alpha})}{I}.
    \end{aligned}
  }
\end{equation}
The orbital angle $\phi$ is slaved to the system through $\dot{\phi} = v_p|\mathbf{P}_0|\sin\Psi/R$ and can be integrated once the solution of the above is known.

We now rewrite Eq.~\eqref{dyn-sys-dimens} in dimensionless form. Let
\begin{equation}
    s \equiv |P_0|, 
    \qquad 
    \lambda \equiv \frac{k}{\xi},
    \qquad 
    g \equiv \frac{\Gamma}{\xi_r}.
\end{equation}
The timescale introduced by confinement is $\lambda^{-1}$, while the lengthscale introduced by active displacement of the cluster is
\begin{equation}
    R_0 \equiv \frac{v_p s}{\lambda}.
\end{equation}
We therefore introduce the dimensionless variables
\begin{equation}
    \tau = \lambda t,
    \qquad
    \rho = \frac{R}{R_0},
\end{equation}
In these variables, the radial equation becomes
\begin{align}
    \frac{d\rho}{d\tau}
    &=
    \frac{1}{\lambda R_0}
    \left(
        v_p s \cos\Psi - \lambda R
    \right) \nonumber \\
    &=
    \frac{v_p s}{\lambda R_0}
    \left(
        \cos\Psi - \rho
    \right) \nonumber \\
    &=
    \cos\Psi - \rho .
\end{align}
Similarly, the equation for $\Psi$ gives
\begin{align}
    \frac{d\Psi}{d\tau}
    &=
    \frac{c}{(1-c)\lambda}
    \left[
        \frac{v_p \bar d}{I}\sin(\theta-\bar\alpha)
        +
        s
        \left(
            g R - \frac{v_p}{R}
        \right)
        \sin\Psi
    \right] \nonumber \\
    &=
    \frac{c v_p \bar d}{I(1-c)\lambda}\sin(\theta-\overline{\alpha})
    +
    \frac{s}{\lambda}
    \left(
        g R_0 \rho - \frac{v_p}{R_0\rho}
    \right)
    \sin\Psi .
\end{align}
Using $R_0=v_p s/\lambda$, the second term simplifies as
\begin{align}
    \frac{s}{\lambda}
    \left(
        g R_0 \rho - \frac{v_p}{R_0\rho}
    \right)
    &=
    \frac{s}{\lambda}
    \left(
        g \frac{v_p s}{\lambda}\rho
        -
        \frac{\lambda}{s}\frac{1}{\rho}
    \right) \nonumber \\
    &=
    \frac{g v_p s^2}{\lambda^2}\rho
    -
    \frac{1}{\rho}.
\end{align}
Thus the dimensionless dynamics is ($\dot X$ now means $\frac{d X}{d\tau}$ with respect to adimensioned time $\tau$) % YES everything checked
\begin{subequations}
\begin{empheq}[box=\fbox]{align}
    \dot{\rho}
    &=
    \cos\Psi-\rho,
    \label{eq:rho_dimless}
    \\
    \dot{\Psi}
    &=
    c\eta\sin\chi
    +
    \left(
        \mu\rho-\frac{1}{\rho}
    \right)
    \sin\Psi,
    \label{eq:Psi_dimless}
    \\
    \dot{\theta}
    &=
    \mu\rho\sin\Psi - (1-c) \eta \sin( \theta - \overline{\alpha}),
    \label{eq:chi_dimless}
\end{empheq}
\end{subequations}
where dots denote derivatives with respect to $\tau$, and where the two independent dimensionless parameters are
\begin{equation}
\boxed{
    \eta
    \equiv
    \frac{v_p\bar d}{I\lambda(1-c)},
    \qquad
    \mu
    \equiv
    \frac{g v_p s^2}{\lambda^2}
    =
    \frac{\Gamma}{\xi_r}
    \frac{v_p |P_0|^2}{(k/\xi)^2}.}
\end{equation}
The orbital angle is slaved to the variables $\rho, \Psi$ through
\begin{equation}
    \dot{\phi}
    =
    \frac{\sin\Psi}{\rho}.
\end{equation}
The parameter $\eta$ measures the strength of the rigid-body rotation induced by the static disorder of the cluster, relative to the confinement rate. The parameter $\mu$ measures the strength of the Marangoni-induced internal polarity rotation relative to the same confinement rate. We will see, as in \cite{Kato2025}, that we need $\mu > 1$ to observe interesting behavior; in turn, we will perform the subsequent analysis perturbatively for small values of $\eta$, allowing us to use the already solved single-colloid case as a base for $\eta=0$ (it is analogous to $\bar d=0$.)

\subsection*{Off-centered, circle-like trajectories for small values of $\eta$}

We now look for the class of trajectories relevant to the observed motion,
which resemble deformed off-centered circles. We thus seek solutions for which
the radial coordinate $\rho$ is periodic in the internal phase, while allowing
the global orbital angle $\phi$ to accumulate a small mismatch over many
cycles. This produces approximately circular off-centered trajectories with
slow precession.

\subsubsection{Fully symmetric polarities: the $\eta=0$ base case}
We first consider the simpler case $\eta=0$, which was already present in the
single-colloid analysis of Ref.~\cite{Kato2025}. Introducing
\begin{equation}
    \chi=\theta-\overline{\alpha},
\end{equation}
we have $\dot\chi=\dot\theta$, since $\overline{\alpha}$ is constant. In the
absence of the active rigid-body moment, the equations for $\rho$ and $\Psi$
decouple from $\chi$:
\begin{subequations}
\begin{align}
    \dot{\rho}
    &=
    \cos\Psi-\rho,
    \\
    \dot{\Psi}
    &=
    \left(
        \mu\rho-\frac{1}{\rho}
    \right)\sin\Psi,
    \\
    \dot{\chi}
    &=
    \mu\rho\sin\Psi.
\end{align}
\end{subequations}
For $\mu>1$, there is an exact rotating solution
\begin{equation}
    \rho_0
    =
    \frac{1}{\sqrt{\mu}}
    =
    \cos\Psi_0.
\end{equation}
For counterclockwise rotation,
\begin{equation}
    \sin\Psi_0
    =
    \frac{\sqrt{\mu-1}}{\sqrt{\mu}}
    >
    0.
\end{equation}
The corresponding internal and orbital angular velocities are
\begin{equation}
    \dot{\chi}
    =
    \mu\rho_0\sin\Psi_0
    =
    \sqrt{\mu-1},
\end{equation}
and
\begin{equation}
    \dot{\phi}
    =
    \frac{\sin\Psi_0}{\rho_0}
    =
    \sqrt{\mu-1}.
\end{equation}
The laboratory-frame trajectory in this case is then a perfect circle of radius
$R_0/\sqrt{\mu}$. 
\subsubsection{Perturbation theory for nonzero $\eta$}
For small $\eta$, this deformation can be computed perturbatively. 
We focus on solutions for which the orbital angle $\phi$ remains monotonic, so that it may be
used instead of time as the independent variable. 
Recall that 
\begin{equation}
\label{eq:phidot-0}
    \dot \phi = \frac{\sin \Psi}{\rho}, 
\end{equation}
so that dividing
Eqs.~\eqref{eq:rho_dimless} and \eqref{eq:Psi_dimless} by
Eq.~\eqref{eq:phidot-0}, we obtain
\begin{subequations}
\begin{align}
    \rho'
    \equiv
    \frac{d\rho}{d\phi}
    &=
    \rho \cot \Psi - \frac{\rho^2}{\sin \Psi},
    \label{eq:rho_phi}
    \\
    \Psi'
    \equiv
    \frac{d\Psi}{d\phi}
    &=
    \mu \rho^2 - 1 + c \eta \rho \frac{\sin \chi}{\sin \Psi}.
    \label{eq:Psi_phi} \\ 
    \chi'
    \equiv
    \frac{d\chi}{d\phi}
    &=
    1 + O(\eta).
    \label{eq:chi_phi}
\end{align}
\end{subequations}
We now expand the solutions to first order in $\eta$ using our solution to the base case $\eta=0$ above
\begin{equation}
    \rho = \frac{1}{\sqrt{\mu}} + \eta \rho_1 + O(\eta^2), \quad \Psi = \arccos \frac{1}{\sqrt{\mu}} + \eta \Psi_1 + O(\eta^2). 
\end{equation}
Developing Eqs\eqref{eq:rho_phi},\eqref{eq:Psi_phi}, and noting that $\chi = \phi + \chi_0 + O(\eta)$ where $\chi_0$ is a constant we can set to $0$ after changing the origin of $\phi$, we obtain 
\begin{subequations}
\begin{align}
    \rho_1'
    &=
    -\frac{\rho_1}{\sqrt{\mu-1}}-\frac{\Psi_1}{\sqrt{\mu}},
    \label{eq:rho_phi_pert}
    \\
    \Psi_1'
    &=
    2\sqrt{\mu} \rho_1 + \frac{c \sin \phi}{\sqrt{\mu-1}}.
    \label{eq:Psi_phi_pert} 
\end{align}
\end{subequations}

To solve the above differential system, we Fourier transform the above $\phi \to \alpha$
\begin{subequations}
\begin{align}
    i \alpha \hat{\rho}_1 
    &=
    -\frac{\hat{\rho}_1}{\sqrt{\mu-1}}-\frac{\hat{\Psi}_1}{\sqrt{\mu}},
    \label{eq:rho_laplace}
    \\
    i \alpha \hat{\Psi}_1
    &=
    2\sqrt{\mu} \hat{\rho}_1 \mp \frac{i c}{2\sqrt{\mu-1}} \delta(\alpha\pm1).
    \label{eq:Psi_laplace} 
\end{align}
\end{subequations}
We see that for $\alpha \neq \pm 1$, the above has a nonzero solution only for \begin{equation*}
    \alpha = \frac{i}{2\sqrt{\mu-1}} (1 \pm \sqrt{9-8 \mu}).
\end{equation*}
This frequency verifies $\text{Re}(i \alpha) <0$ and therefore gives a vanishing contribution to $\rho_1, \Psi_1$ in the long-time limit.

However, for $\alpha = \pm 1$, one has an oscillating solution which corresponds to the long-time behavior of the trajectory. Performing the algebra, this implies the following long-time behavior for $\rho_1$ and $\Psi_1$
\begin{align}
    \rho_1(\phi) &\approx \frac{c}{\mu^{3/2}} (\cos \phi - \sqrt{\mu-1} \sin \phi), \\ \Psi_1(\phi) &\approx \frac{c}{\mu} \left(2\sin \phi  + \frac{\mu-2}{\sqrt{\mu-1}} \cos \phi \right).
\end{align}
After a shift of the orbital angle $\phi \to \phi - \phi_1$ with $\phi_1 \equiv - \arccos \frac{1}{\sqrt{\mu}}$, and using 
\[
\sin \left(-\arccos \frac{1}{\sqrt{\mu}}\right) = -\sqrt{\frac{\mu-1}{\mu}},\]
one can write 
\begin{equation}
    \rho_1(\phi) \approx \frac{c}{\mu} \cos(\phi-\phi_1),
\end{equation}
which is the result in the main text. As for $\Psi_1$, one has 
\begin{equation}
    \Psi_1(\phi) \approx \frac{c}{\sqrt{\mu}} \left(\sin(\phi-\phi_1) - \frac{\cos(\phi-\phi_1)}{\sqrt{\mu-1}} \right)
\end{equation}

\subsubsection{Raft velocity}
We finally compute the absolute velocity of the raft. In reduced units,
\begin{equation}
    v_{\rm raft}^2
    =
    \dot{\rho}^2+\rho^2\dot{\phi}^2.
\end{equation}
Using the exact identities
\begin{equation}
    \dot{\rho}
    =
    \cos\Psi-\rho,
    \qquad
    \dot{\phi}
    =
    \frac{\sin\Psi}{\rho},
\end{equation}
we obtain
\begin{equation}
    v_{\rm raft}^2
    =
    \left(
        \cos\Psi-\rho
    \right)^2
    +
    \sin^2\Psi.
\end{equation}
Since
\begin{equation}
    \cos\Psi_0=\rho_0=\frac{1}{\sqrt{\mu}},
\end{equation}
the first term is of order $O(\eta^2)$, and therefore
\begin{equation}
    v_{\rm raft}^2
    =
    \sin^2\Psi
    +
    O(\eta^2).
\end{equation}
Using 
\begin{equation*}
    \sin^2 \Psi = \sin^2 \Psi_0 + 2 \eta \Psi_1 \sin (\Psi_0) \cos (\Psi_0) + O(\eta^2) = \frac{\mu-1}{\mu} + 2 \eta \Psi_1 \frac{\sqrt{\mu-1}}{\mu}+ O(\eta^2),
\end{equation*}
the first-order expansion is
\begin{equation}
    v_{\rm raft}^2
    =
    \frac{\mu-1}{\mu}
    +
    \frac{2c\eta}{\mu^{3/2}}
    \left[
        \sqrt{\mu-1} \sin(\phi-\phi_1) - \cos(\phi-\phi_1)
    \right]
    +
    O(\eta^2).
    \label{eq:speed_squared_perturbative}
\end{equation}
Equivalently,
\begin{equation}
    \boxed{
    v_{\rm raft}
    =
    \sqrt{\frac{\mu-1}{\mu}}
    +
    \frac{c\eta}{\mu\sqrt{\mu-1}}
    \left[
        -\cos (\phi-\phi_1)
        +
        \sqrt{\mu-1}\,\sin (\phi-\phi_1)
    \right]
    +
    O(\eta^2).
    }
    \label{eq:speed_perturbative}
\end{equation}

For $c>0$, the raft speed is maximal when
\begin{equation}
    \phi-\phi_1
    =
    \pi-\arctan\sqrt{\mu-1}
    \qquad [2\pi],
\end{equation}
and its maximal value is
\begin{equation}
    v_{\rm raft}^{\rm max}
    =
    \sqrt{\frac{\mu-1}{\mu}}
    +
    \frac{c\eta}
    {\sqrt{\mu(\mu-1)}}
    +
    O(\eta^2).
\end{equation}
It is minimal when
\begin{equation}
    \phi-\phi_1
    =
    -\arctan\sqrt{\mu-1}
    \qquad [2\pi],
\end{equation}
with
\begin{equation}
    v_{\rm raft}^{\rm min}
    =
    \sqrt{\frac{\mu-1}{\mu}}
    -
    \frac{c\eta}
    {\sqrt{\mu(\mu-1)}}
    +
    O(\eta^2).
\end{equation}

By contrast, the raft is furthest from the droplet center, and hence closest
to the contact line, when
\begin{equation}
    \phi-\phi_1=0
    \qquad [2\pi].
\end{equation}
At this phase,
\begin{equation}
    v_{\rm raft}
    =
    \sqrt{\frac{\mu-1}{\mu}}
    -
    \frac{c\eta}
    {\mu\sqrt{\mu-1}}
    +
    O(\eta^2),
\end{equation}
so the raft moves more slowly than its mean speed while producing its
strongest contact-line deformation. The model therefore predicts a finite
phase shift between the extrema of the raft speed and of the radial
deformation. For large $\mu$, this phase shift approaches one quarter of an
orbital period as $\arctan(\sqrt{\mu-1}) \to \frac{\pi}{2}$.

\subsection*{Numerical simulations}
For simulations, we use a non-harmonic potential 
in order to see how robust our predictions above are to the choice of the specific potential:
\begin{equation}
 U(r)=\frac{g_0}{2}r^2+\frac{A}{6}\left[(r^2-s_0)^3+s_0^3\right].
 \label{eq:Vselected}
\end{equation}
For $A>0$, $U'(r)=rg(r)>0$ for every $r>0$: the origin is the unique minimum and the force is always inward. The nonlinearity only makes the radial stiffness vary over the spatial extent of the raft.

\begin{table}[t]
\caption{\label{tab:geometry}Body-frame geometry and polarity offsets of the representative V raft.}
\begin{ruledtabular}
\begin{tabular}{crrrrr}
$i$ & $q_{i,x}$ & $q_{i,y}$ & $d_i$ & $\alpha_i$ & $\delta_i$\\
1&0&$-0.030841$&0.030841&$-1.570796$&$-0.027687$\\
2&$-0.014838$&$-0.005140$&0.015704&$-2.808119$&0.008166\\
3&$-0.029677$&0.020561&0.036102&2.535702&$-0.057223$\\
4&0.014838&$-0.005140$&0.015704&$-0.333473$&0.159725\\
5&0.029677&0.020561&0.036102&0.605891&$-0.082980$\\
\end{tabular}
\end{ruledtabular}
\end{table}
The remaining parameters are
\begin{equation}
A=12.8330,
 \quad \mu=2.12827,\quad c=0.662437.
 \label{eq:params}
\end{equation}
One then computes $\eta$ from the given raft geometry, which varies as specified below.

The equations were integrated with fourth-order Runge--Kutta. The representative run extends to $t=1800$ with $\Delta t=0.003$; the first $70\%$ is discarded and the final 100 complete orbital turns are analyzed. We note the cycle center displacement by $D$.

\subsubsection{Decomposition of the torques on the solid raft}

\begin{figure}[t]
 \includegraphics[width=.98\textwidth]{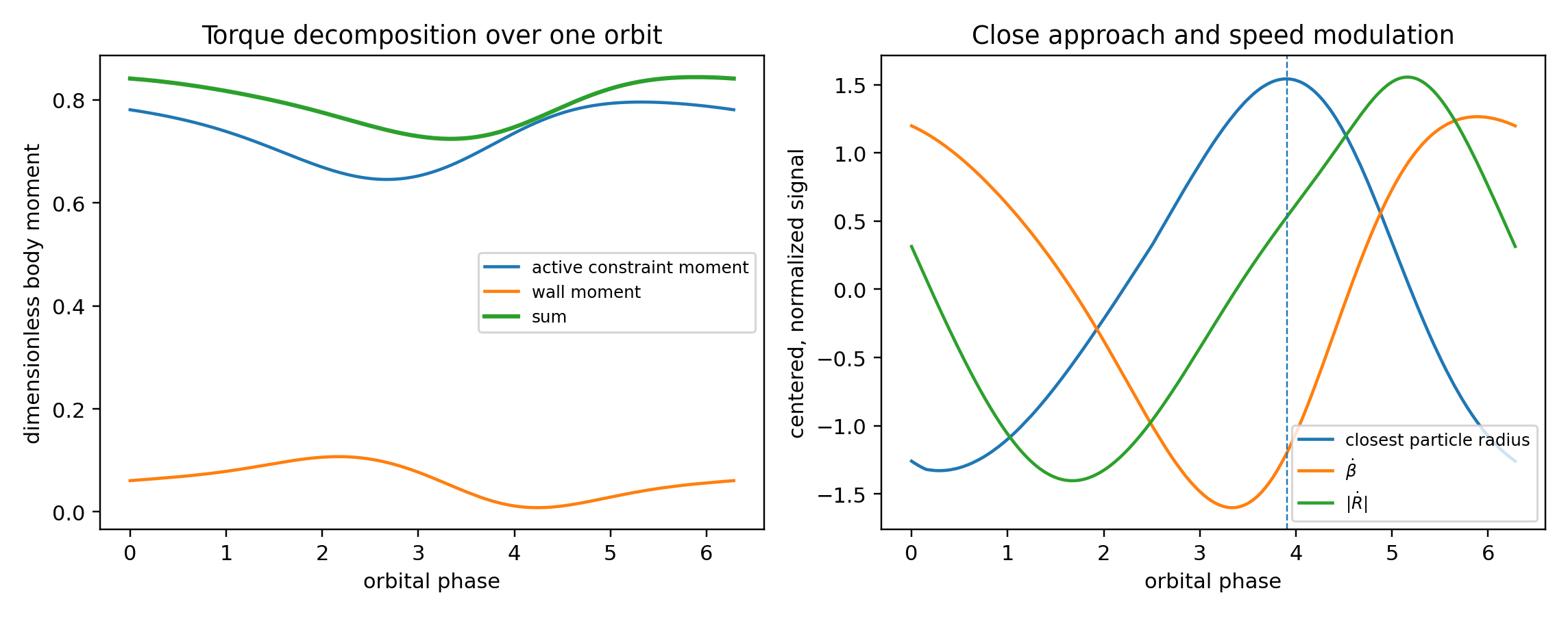}
 \caption{\label{fig:torques}Mechanical signals during one late orbit. Left: active constraint moment, nonlinear-wall moment, and their sum. Right: closest-particle radius, rigid-body angular speed, and center speed, centered and normalized separately. The dashed line marks closest approach. We see that raft speed is quite low at this point.}
\end{figure}

Figure~\ref{fig:torques} decomposes the torque contributions over one cycle. The cycle-averaged self-propelling and potential $U$ contributions inside the brackets are represented Fig.~\ref{fig:torques}. We see the self-propulsion of individual JPs coupled with the solid geometry supplies most of the mean rigid-body rotation $\dot \beta$, while the nonlinear-wall moment provides a small correction. This coupling permits a strongly off-centered orbit along with a rotating raft.

The rigid-body angular speed is anticorrelated with the radius of the closest particle, as seen Fig.~\ref{fig:torques}. 
The raft therefore rotates more slowly when one side approaches the boundary.

\subsubsection{Geometry and polarity}

We now test the model on a set of explicit raft geometries. The confining potential and all reduced dynamical parameters in Eq.~\eqref{eq:params} are held fixed. Every simulation starts from the same center position and the same initial collective-polarity phase.

\begin{figure}[t]
 \includegraphics[width=.99\textwidth]{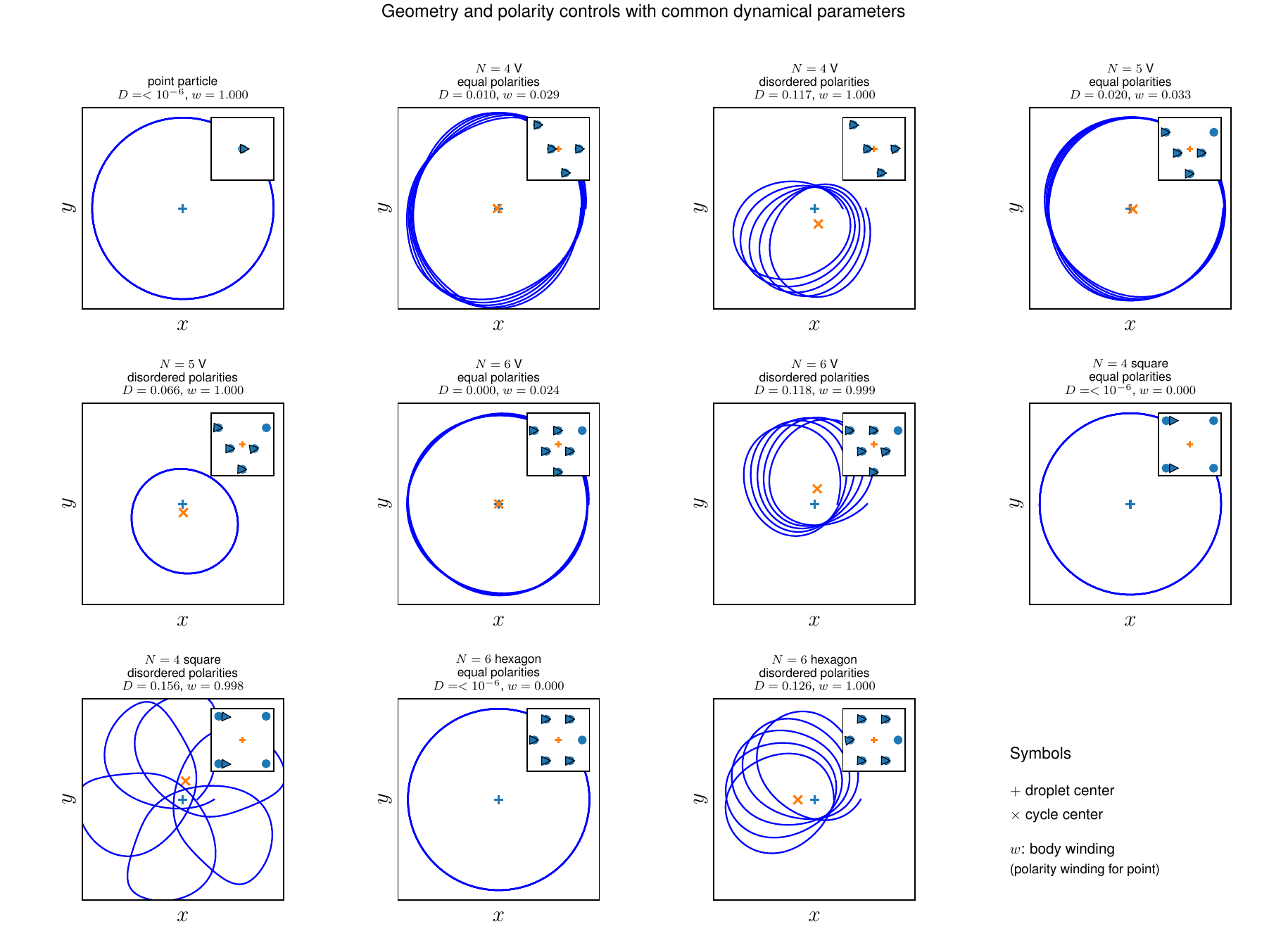}
 \caption{\label{fig:geometry-controls}Geometry and polarity controls with common dynamical parameters. Each panel shows one late center-of-mass orbit; the inset gives the actual body-frame particle positions and polarity offsets used in that simulation. The plus sign marks the droplet center and the cross the geometrical cycle center. In the panel headings, $w$ is the rigid-body winding for extended rafts and the polarity winding for the point particle.}
\end{figure}

The V-shaped geometries (see Fig.~\ref{fig:geometry-controls} use the same nearest-neighbor spacing $\ell=0.0296768$. Before subtraction of the center of mass, the five-particle V raft reference is
\begin{equation}
 {\cal S}_5=\ell\left\{(0,0),\left(-\frac12,\frac{\sqrt3}{2}\right),(-1,\sqrt3),
 \left(\frac12,\frac{\sqrt3}{2}\right),(1,\sqrt3)\right\}.
 \label{eq:S5compact}
\end{equation}
The $N=4$ V is obtained by removing the right outer particle from the above reference geometry, and the $N=6$ V by adding the rear central site $(0,\sqrt3)\ell$. To control how the absence of a preferred solid axis changes the trajectory, we also check squares and hexagons: the square has side length $\ell$, and the regular hexagon has side length $\ell$. 

For every geometry we compare identical body-frame polarity angles, $\delta_i=0$, with frozen disordered polarities. The V-raft offsets are
\begin{align}
 \bm\delta^{(4)}&=(-0.04843,-0.01258,0.13898,-0.07797),\\
 \bm\delta^{(5)}&=(-0.02769,0.00817,-0.05722,0.15972,-0.08298),\\
 \bm\delta^{(6)}&=(-0.01600,0.01985,-0.04554,0.17141,-0.07130,-0.05842).
 \label{eq:control-offsets}
\end{align}
To isolate the effect of positional geometry, the square uses exactly $\bm\delta^{(4)}$ and the hexagon exactly $\bm\delta^{(6)}$. Thus the square and $N=4$ V share the same polarity pattern, as do the hexagon and $N=6$ V.

Figure~\ref{fig:geometry-controls} displays the actual body-frame geometry and one late orbit for all cases. The quantitative results are listed in Table~\ref{tab:geometry-controls}. The point particle has no rigid-body orientation and remains centered, while its polarity winds once around the droplet center. When all body-frame polarity angles are identical, the active constraint moment vanishes exactly, and the raft essentially remains centered (see Fig.~\ref{fig:geometry-controls}).

\begin{table}[t]
\caption{\label{tab:geometry-controls}Point, V-raft, square, and hexagon controls. Precession is reported for the disordered body-locked trajectories; it is omitted for centered states and for equal-polarity V trajectories that do not body-lock. All quantities are evaluated over the final 80 complete orbital turns.}
\small
\begin{ruledtabular}
\begin{tabular}{llccccc}
geometry & polarities & $P_0$ & $\eta$ & $D$ & $\Delta\varphi_{\rm prec}$ & winding\\
point particle & -- & 1 & 0 & $1.9\times10^{-9}$ & -- & $\Delta\psi/(2\pi)=1.000000$\\
$N=4$ V & equal & 1 & 0 & 0.01024 & -- & $\Delta\beta/(2\pi)=0.02865$\\
$N=4$ V & disordered & 0.99652 & 5.24461 & 0.11727 & $-0.24548$ & $\Delta\beta/(2\pi)=0.99974$\\
$N=5$ V & equal & 1 & 0 & 0.01993 & -- & $\Delta\beta/(2\pi)=0.03305$\\
$N=5$ V & disordered & 0.99636 & 2.32090 & 0.06570 & $2.33\times10^{-5}$ & $\Delta\beta/(2\pi)=1.000001$\\
$N=6$ V & equal & 1 & 0 & $2.36\times10^{-4}$ & -- & $\Delta\beta/(2\pi)=0.02364$\\
$N=6$ V & disordered & 0.99662 & 2.81306 & 0.11753 & 0.24333 & $\Delta\beta/(2\pi)=0.99940$\\
$N=4$ square & equal & 1 & 0 & $7.1\times10^{-9}$ & -- & $\Delta\beta/(2\pi)=0$\\
$N=4$ square & disordered & 0.99652 & 6.57536 & 0.15650 & $-1.38997$ & $\Delta\beta/(2\pi)=0.99783$\\
$N=6$ hexagon & equal & 1 & 0 & $6.3\times10^{-9}$ & -- & $\Delta\beta/(2\pi)=0$\\
$N=6$ hexagon & disordered & 0.99662 & 2.78963 & 0.12559 & $-0.39301$ & $\Delta\beta/(2\pi)=1.00002$\\
\end{tabular}
\end{ruledtabular}
\end{table}

\newpage

% ============================================================
\section{Comparison between the measured droplet velocity
and the deformation-based prediction (for Fig.4)}
\label{sec:velocity_comparison}
% ============================================================

To compare the experimentally measured translational motion of
the droplet with the velocity predicted from its deformation, we
analyzed the droplet contour and centroid independently at each
time point. The theoretical prediction therefore uses only the
measured deformation of the droplet as an input and does not use
the measured translational velocity as a fitting parameter.

From each frame of the high-speed microscopy recordings, we
determined the position of the droplet center, $\vect{x}^{D}_{c}(t)$ from the tracked droplet contour. The contour was then expressed
in polar coordinates with respect to the center which was taken from a smoothed trajectory by the raft period. The contour was fitted with the Fourier components up to $n=5$:
\begin{equation}
    R(\theta,t)
    =
    a_0(t)
    +
    \sum_{n=2,\dots, 5}
    \left[
        a_n(t)\cos(n\theta)
        +
        b_n(t)\sin(n\theta)
    \right].
    \label{eq:experimental_fourier}
\end{equation}
Here, $a_0(t)$ represents the instantaneous mean radius of the
droplet, while $a_n(t)$ and $b_n(t)$ describe the deformation of
the contact line. The $n=1$ contribution corresponds to a
translation of the origin rather than to a genuine shape
deformation and is not used in the deformation-based prediction.
The non-dimensional complex Fourier
coefficients ($n\geq 2$)
\begin{equation}
    \alpha_n(t) =\frac{a_n(t) - i b_n(t)}{2a_0},
    \label{eq:complex_exp_modes}
\end{equation}
consistently with that in the main paper by setting $\varepsilon=1$.

Both the droplet-center trajectory and the deformation modes were
processed using the same temporal smoothing procedure. This is
particularly important because both the experimental velocity and
the theoretical prediction involve time derivatives.

The time series $x_c^D(t),\quad y_c^D(t),\quad a_0(t),\quad a_n(t),\quad b_n(t)$ were smoothed using a Savitzky--Golay filter with a third-order polynomial and a temporal window of $\Delta t_{\mathrm{SG}} = 0.031~\mathrm{s}$.

The deformation-induced translational velocity $\vect{U}_\mathrm{geom}$ derived in
Eq.~(3) in the main text can be evaluated directly from the experimentally
measured Fourier coefficients $\alpha_n(t)$. 

\begin{figure}[t]
    \centering
    \includegraphics[
        width=0.48\textwidth
    ]{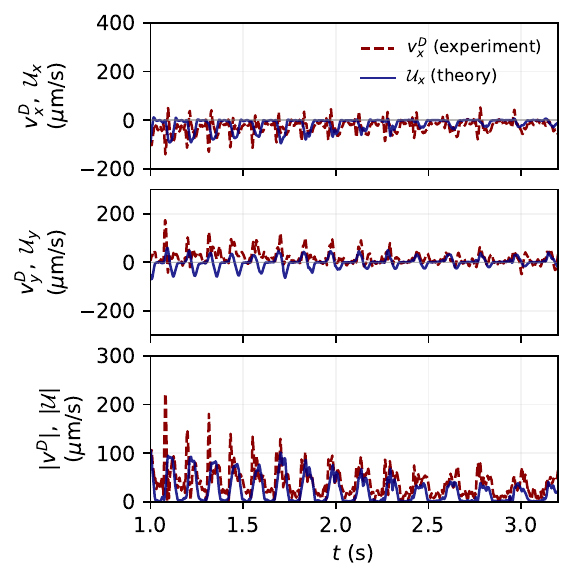}

    \caption{
    \textbf{Time-resolved comparison of the measured and
    deformation-predicted droplet velocities for sample 2.}
    Top and middle panels show the $x$ and $y$ components,
    respectively, of the experimentally measured droplet velocity
    $\bm{v}^{D}$ (red dashed lines) and the deformation-based
    prediction $\bm{\mathcal U}$ (blue solid lines).
    The bottom panel shows the corresponding speed magnitudes
    $|v^{D}|$ and $|\mathcal U|$.
    The theoretical velocity is calculated directly from the
    experimentally measured contact-line deformation modes without
    an adjustable multiplicative prefactor.
    }
    \label{fig:velocity_components_sample2}
\end{figure}

As shown in Fig.~\ref{fig:velocity_components_sample2}, some discrepancy remains in the individual velocity components, indicating imperfect agreement in the instantaneous direction of motion. In contrast, the predicted speed reproduces the temporal modulation of the measured speed more closely.

% \section{Comparison between the measured droplet velocity components
% and the deformation-based prediction}
% \label{sec:velocity_comparison_more}

% By the deformation-based prediction,
% each component of velocity is calculated from the experimentally measured Fourier modes of the
% droplet contour using Eq.~(3). Importantly,
% the measured centroid velocity is not used as an input to this
% calculation, and no multiplicative rescaling of
% ${\mathcal U}$ is introduced in the comparison.

% Figure~\ref{fig:velocity_components_sample2} shows a representative
% example for the sample in the main text. The $x$ and $y$ components of the measured
% velocity are compared directly with the corresponding components of
% the deformation-based prediction. The bottom panel shows the
% corresponding speed magnitudes.

% The predicted velocity is generally smaller in magnitude than the
% measured velocity. Nevertheless, the theoretical curves reproduce
% part of the temporal variation of the experimental velocity,
% including repeated increases and decreases in both the individual
% velocity components and the speed magnitude. This indicates that
% the non-reciprocal deformation of the droplet boundary contributes
% to the observed translational dynamics, although it does not by
% itself account for the full velocity amplitude.

% ============================================================

\end{document}